\documentclass{optica-article}

\journal{opticajournal} 

\articletype{Research Article}

\usepackage{lineno}
\usepackage{soul}
\usepackage[english=usenglishmax]{hyphsubst}
\begin{document}

\title{Do different kinds of photon-pair sources have the same indistinguishability in quantum silicon photonics?}

\author{Jong-Moo Lee,\authormark{1,*} Alessio Baldazzi,\authormark{2,**} Matteo Sanna,\authormark{2} Stefano Azzini,\authormark{2} Joon Tae Ahn,\authormark{1}  Myung Lae Lee,\authormark{1} Young-Ik Sohn,\authormark{3} and Lorenzo Pavesi\authormark{2}}

\address{\authormark{1} ETRI, 218 Gajeong-ro, Daejeon, 34129, South Korea\\
\authormark{2} Department of Physics, University of Trento, via Sommarive 14, Trento, 38123, Italy\\
\authormark{3} KAIST, 291 Daehak-ro, Daejeon, 34141, South Korea}

\email{\authormark{*}jongmool@etri.re.kr, \authormark{**}alessio.baldazzi@unitn.it} 



\begin{abstract}
In the same silicon photonic integrated circuit, we compare two types of integrated degenerate photon-pair sources (microring resonators or waveguides) by means of Hong-Ou-Mandel (HOM) interference experiments. Two nominally identical microring resonators are coupled to two nominally identical waveguides which form the arms of a Mach-Zehnder interferometer. This is pumped by two lasers at two different wavelengths to generate, by spontaneous four-wave mixing, degenerate photon pairs. In particular, the microring resonators can be thermally tuned in or out of resonance with the pump wavelengths, thus choosing either the microring resonators or the waveguides as photon-pair sources, respectively. In this way, an on-chip HOM visibility of 94\% with microring resonators and 99\% with straight waveguides is measured upon filtering. We compare our experimental results with theoretical simulations of the joint spectral intensity and the purity of the degenerate photon pairs. We verify that the visibility is connected to the sources' indistinguishability, which can be quantified by the overlap between the joint spectral amplitudes (JSA) of the photon pairs generated by the two sources. We estimate a JSAs overlap of 98\% with waveguides and 89\% with microring resonators.
\end{abstract}

\section{Introduction}
\label{sec:intro}

In quantum computing, it is important to distinguish between physical and logical qubits. The firsts are the two-level physical systems that are processed in the quantum hardware, while the seconds are clusters of redundant physical qubits which store the information \cite{Shaw_2008,Viola_2001,Heeres_2017,Rist__2015,Kapit_2016}.
On photonic platforms, the number of physical qubits is typically associated with the number of photons, whose quality is intrinsically quantified by the purity of the photon states and extrinsically by their indistinguishability \cite{indi_pure_photo22,indi_photo22}.
The demand for reliable physical qubits translates into engineering high-quality sources of photons. Currently, there are two main types of integrated sources: deterministic and probabilistic. In the case of deterministic sources, the integration in a  silicon photonic-integrated circuit (SiPIC) is still very demanding because of the high coupling losses and the complex fabrication process. On the other side, there have been many reports of probabilistic photon-pair sources integrated in SiPIC over the last 10 years \cite{Azzini2012,Engin2013}. Then, the correlation between the photons is exploited to produce single photon sources through the heralding mechanism, where one photon of the pair is detected to herald the presence of its twin photon \cite{Signorini2020on-chip}. 
The quality of a generated single photon state is quantified by how close to zero is the second order coherence function $g^{(2)}(0)$ in the case of deterministic sources, or how close to zero is the heralded second order coherence function $g^{(2)}_h(0)$ in the case of probabilistic ones.

In this paper, we aim to address the following question: what type of source does perform better among the probabilistic sources? This is not a trivial question. The limited amount of physical qubits in the so-called noisy intermediate scale quantum (NISQ) computing era prevents the use of error-correction techniques in universal quantum computing hardware and brings people to work with quantum simulators, i.e. devices that perform non-universal and problem-focused algorithms \cite{Feynman82,johnson2014quantum}. Therefore, the most suitable source of photons should be determined by the requirements of the specific implemented algorithm. The ideal close-term aim would be the realization of a sources' library, which can be consulted as needed and updated with the improvements of the sources present in the list. 

Among probabilistic sources, the nonlinearity of the photonic material is typically used to create pairs of photons through nonlinear spontaneous parametric processes. Once the generation is achieved, the generated photons are led to reconfigurable networks of Mach-Zehnder interferometers (MZIs) to produce interference patterns. These are used to qualify the quality of the generated photons \cite{Signorini2020on-chip}. 
Photon-pair generation and interference between the photons within a SiPIC have gained more and more attention with the growing anticipation toward a fault-tolerant quantum computer based on linear optical quantum computation \cite{raussendorf2001a, briege2009measurement-based, bartolucci2021fusion-based, bombin2021modular, viglia2021error-protected}. In silicon photonics, photon pairs can be generated by spontaneous four-wave mixing (SFWM) in a long waveguide \cite{silverstone2014on-chip, lee2019noise} or a compact microring resonator \cite{farugue2018on-chip, llewellyn2020chip-to-chip}. Quantum interference of photons in a SiPIC has been demonstrated using heralded photons from non-degenerate photon pairs \cite{farugue2018on-chip, llewellyn2020chip-to-chip} or using degenerate photon pairs composed of two photons hard to distinguish from each other \cite{silverstone2014on-chip}. The photon pairs degenerate in the signal-idler energies can be generated by pumping the SFWM process with two laser wavelengths in, e.g., spiraled waveguides which form the arms of an integrated MZI. The propagation path of the photons can be controlled by adjusting the phase of the MZI, and on-chip quantum interference such as the Hong-Ou-Mandel (HOM) effect can be observed \cite{silverstone2014on-chip, lee2022controlled-not}.
On the other side, using integrated microring resonators as photon-pair sources, the HOM effect \cite{hong1987measurement,branczyk2017hong,bouchard2020two} of heralded photons has also been reported more recently \cite{farugue2018on-chip, llewellyn2020chip-to-chip}, and a degenerate photon-pair generation from a single microring resonator has been demonstrated \cite{he2015ultracompact}. However, on-chip quantum interference of degenerate photon pairs from two microring resonators has not been shown yet, and neither its direct comparison with an experiment using waveguide spirals. 

Here, we aim at comparing probabilistic degenerate photon-pair sources made of microring resonators or waveguides. In order to make a fair comparison, we used sources integrated in the same SiPIC. By using microring resonators coupled to waveguides that form the arms of an MZI, we measure the on-chip HOM interference to assess the quality of the generated photon pairs. Microring resonators can be thermally tuned in or out of resonance with the pump wavelengths, to turn on and off the microring-based photon sources, respectively. In this way, we can directly compare the on-chip interference of the degenerate photon pairs from the microrings with respect to the one from the waveguides.
Our configuration, composed of the degenerate SFWM with two sources put in parallel inside an MZI, allows us to have a configuration not limited by the purity but by the indistinguishability of the sources.

The paper is organized in this way. In section \ref{sec:exp_setup} we describe the fabricated devices and the used experimental setups. Section \ref{sec:nonlin} reports the measurement results for the photon-pair sources. Section \ref{sec:pair} shows the results of quantum interference experiments for waveguide-based sources. In section \ref{sec:ring}, the results of quantum interference experiments on waveguide- and microring-resonator-based photon-pair sources in the same integrated photonic circuit are presented. Section \ref{sec:disc} is about a discussion of the characteristic parameters for the different types of photon-pair sources by using both experimental data and simulations. Section \ref{sec:conc} aims at making a final statement about the main results of this paper. Finally, the Appendix details the theoretical analysis and the numerical calculations.

\section{Quantum silicon PICs and experimental setups}
\label{sec:exp_setup}

The generation of photon pairs in a SiPIC can be achieved through SFWM processes in a low squeezing regime \cite{Quesada:22,Wright:2018,Parmigiani:2019,Cordier:19}. Typically, two pump photons (at wavelengths $\lambda_{P1}$ and $\lambda_{P2}$) are converted into two newly generated idler and signal photons (at wavelengths $\lambda_i$ and $\lambda_s$) \cite{Signorini2020on-chip}. Non-degenerate SFWM is achieved when identical pump photons wavelength ($\lambda_{P1}= \lambda_{P2}$) are used so that the generated photons have different wavelengths \cite{Fukuda:05,Melloni:08,Friis:18,signorini2021silicon}, while in degenerate SFWM the situation is inverted, i.e. different pump photon wavelengths ($\lambda_{P1}\neq \lambda_{P2}$) and equal generated photons wavelength ($\lambda_i= \lambda_s= 2(\lambda_{P1}^{-1}+\lambda_{P2}^{-1})^{-1}$)\cite{SchmittbergerMarlow:20}. 
In both cases, the state of the output photons is in general a squeezed state: the two-mode squeezed state for the non-degenerate case, and the single-mode squeezed state for the degenerate one \cite{Lvovsky:14,Bagchi_2020}. Both states have useful properties for quantum applications. For example, the non-degenerate case is used in combination with the heralding procedure to obtain probabilistic single photon sources, while the degenerate case produces a couple of correlated identical single photons. In this work, the latter case has been implemented in a SiPIC, where the generated pairs of degenerate photons are used to obtain HOM interference.

\begin{figure}[http]
    \centering
    \includegraphics[scale=0.79]{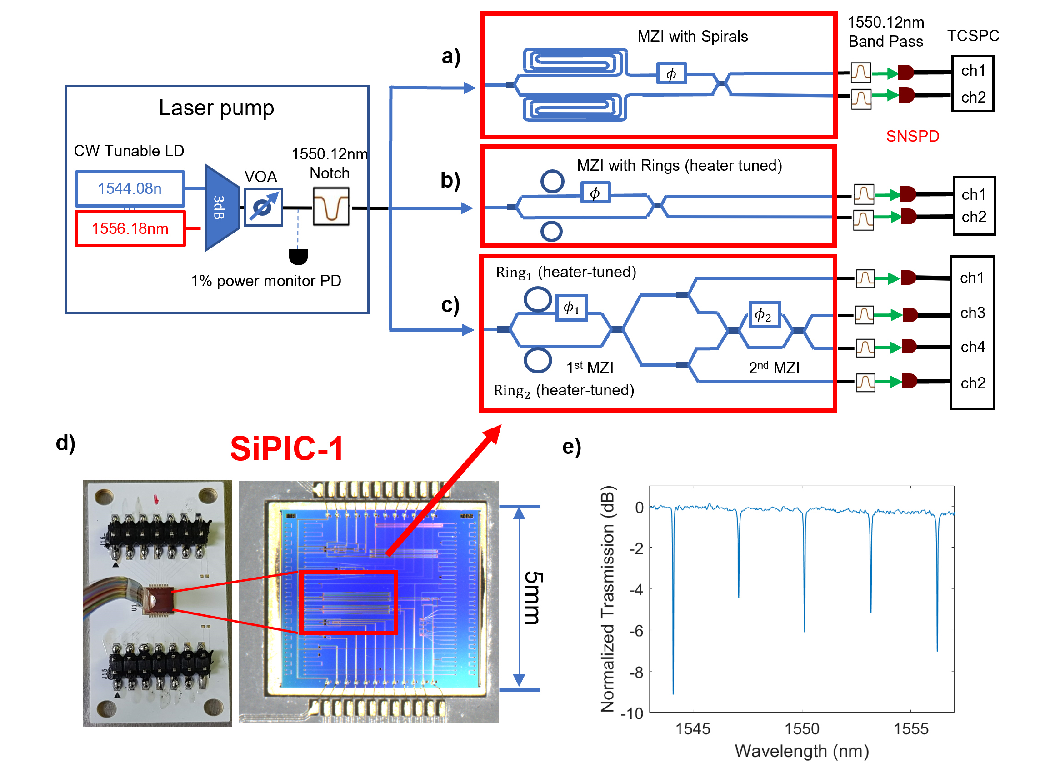}
\caption{\small{(top) Experimental set-up to measure photon-pair generation and multi-source quantum interference in SiPIC-1. The blue rectangle contains the pump lasers apparatus. The thick blue lines with arrows represent the fibers coupling the pump beams to the SiPIC, represented by the different circuits enclosed by the red rectangles. The circuit in (a) represents the photon-pair sources based on spiral waveguides that form the two arms of an MZI. The circuit in (b) represents photon-pair sources formed by microrings. The circuit in (c) represents composite photon-pair sources based on both waveguides and microrings. These are followed by a second MZI to measure the quantum interference of the generated photons. On the right, the detection channels (ch1-ch4) are represented. These are based on a sequence of optical fibers, band-pass filters, superconductor nanowire single photon detectors (SNSPDs) and a time-correlated single photon counting module interfaced to a computer for further processing. (d) On the left, photograph of the packaged SiPIC-1 chip and, on the right, zoomed image of the chip with the circuit highlighted by the red rectangle. (e) Normalized transmission spectrum of the circuit in (b) when the two microring resonators are tuned in resonance with the pump photons wavelengths. 
}} 
    \label{fig:SiPIC-1}
\end{figure}

\begin{figure}[http]
    \centering
    \includegraphics[scale=0.7]{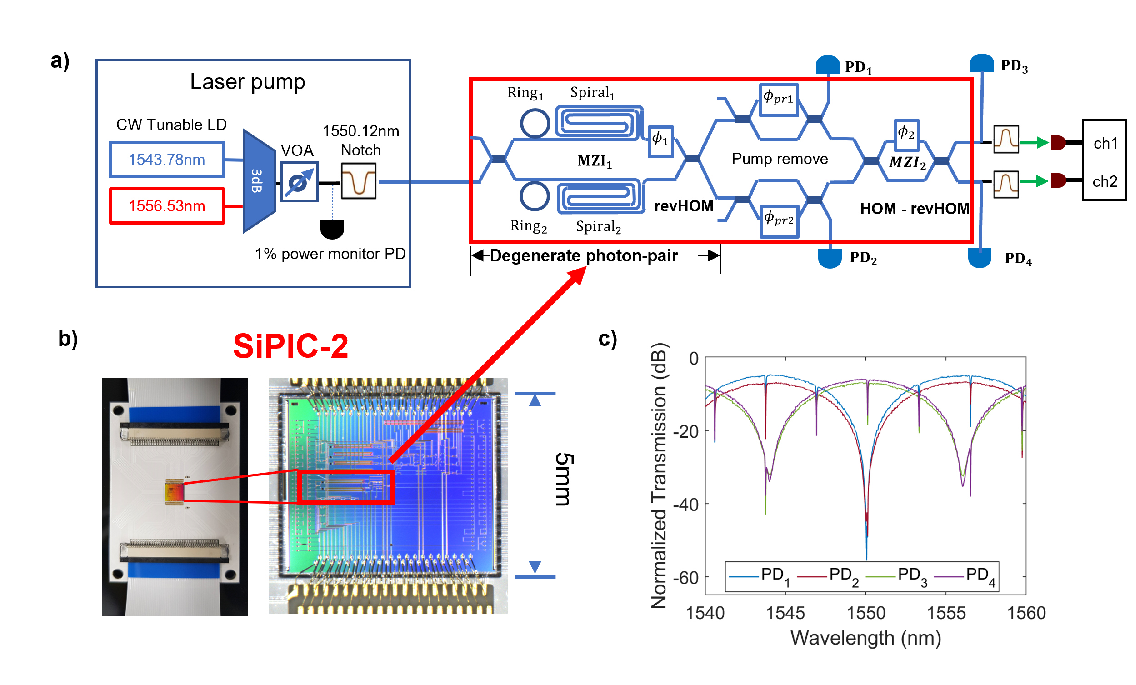} 
\caption{\small{Experimental set-up to measure photon-pair generation and multi-source quantum interference in SiPIC-2. The blue rectangle contains the pump lasers apparatus. The red rectangle contains a scheme of the SiPIC. Four photodiodes (PD$_1$-PD$_4$) are interfaced to the output of SiPIC-2 by optical fibers and grating couplers. In addition, two single photon counting channels (formed by the same sequence as in Fig. \ref{fig:SiPIC-1}) are used to measure the output coincidence counts. (b) On the left, photograph of the packaged SiPIC-2 chip and, on the right, zoomed image of the chip with the circuit highlighted by the red rectangle. (c) Normalized transmission spectra measured by the different photodiodes (blu line PD$_1$, red line PD$_2$, green line PD$_3$, and violet line PD$_4$) while scanning the wavelength of one of the tunable laser diodes.}}
    \label{fig:SiPIC-2}
\end{figure}

Two SiPICs are used in this work. Figs. \ref{fig:SiPIC-1} and \ref{fig:SiPIC-2} show the schematic diagrams of the integrated circuits for the two devices named SiPIC-1 and SiPIC-2, respectively.
Pictures of the actual chips together with their packaging modules are shown in Figs. \ref{fig:SiPIC-1}(d) and \ref{fig:SiPIC-2}(b). The devices are packaged on a metal-based printed-circuit board (MPCB), with 24-port electrical wiring, and are pig-tailed to a 24-channel fiber array. The MPCB is in contact with a thermo-electric cooler (TEC) to control the chip temperature. SiPIC-1 and SiPIC-2 are based on silicon waveguides with a nominal 450 x 220 nm$^2$ cross-section (typically 480 x 210 nm$^2$ after fabrication) and were fabricated through IMEC/Europractice using their passive+ Silicon-on-Insulator (SOI) platform, similarly to our previous report \cite{lee2022controlled-not}. 
Dispersion tailoring was not used to optimize SFWM processes since previous experiments \cite{lee2019noise} show a quite broad generation spectrum for photon pairs in silicon waveguides. 
The measured waveguide propagation loss is 2 dB/cm, a relevant feature for the sources we want to compare. For microring resonators, it determines the quality of the cavity, whose design has the coupler parameters adjusted to the losses. For waveguide spirals, it gives an effective length $L_{\rm eff}=[1-\exp(-\alpha L)]/\alpha$, with $L$ the geometrical length and $\alpha$ the propagation loss per unit of length, which enters quadratically in the SFWM generation rate \cite{helt2012does}. From the analysis of this quantity, we choose the geometrical length of the long spirals in SiPIC-1 and SiPIC-2, finding 1.5 cm as an optimal value.

Fig. \ref{fig:SiPIC-1} top shows schematically the experimental setup based on SiPIC-1. Two continuous wave (CW) tunable laser diodes (CoBrite from IDPhotonics) at $\lambda_{P1}$=1544.08 nm and  $\lambda_{P2}$=1556.18 nm are combined by a 3-dB fiber-optic coupler and provide the pump photons. The combined beam is passed through an optical notch filter (NF) \cite{lee2022controlled-not, lee2019noise} to eliminate photon noise within 1.6 nm bandwidth around $\lambda_i = \lambda_s = 1550.12$ nm, which is the wavelength of the on-chip generated photon pairs. Then, the pump photons are inserted into the selected input fiber of the fiber array which is coupled to the chip by grating couplers, whose coupling loss is measured to be 4.2 dB. In SiPIC-1, different circuits are present (Fig. \ref{fig:SiPIC-1}(a), (b), and (c)). These are based on MZIs with two nominally identical photon-pair sources located in their arms. 
The photon-pair sources are based on 15 mm-long waveguide spirals (Fig. \ref{fig:SiPIC-1}(a)) or on 30 $\mu$m-radius microring resonators (Figs. \ref{fig:SiPIC-1}(b) and (c)). The microrings are coupled to micro-heaters (metal wires on top of the microring) which allow thermal tuning of their resonances. Fig. \ref{fig:SiPIC-1}(e) shows the normalized transmission spectrum of one of the microrings taken by scanning the wavelength of one of the tunable laser diodes: transmission resonances are observed at 1544.08, 1550.12, and 1556.18 nm (i.e. at the wavelengths of the pump photons and of the generated photon-pairs), their free spectral range (FSR) is about 3 nm and their loaded quality factor (Q-factor) is about 1.5x10${^4}$. Micro-heaters are also integrated on one arm of the MZI to act as phase shifters ($\phi$) in order to compensate unwanted phase differences between the two arms. The efficiency of the micro-heater (0.6 x 40 $\mu$m) of the MZI is measured to be about 25 mW/$\pi$ with an overall resistance of about 50 $\Omega$. The efficiency of the heater is relatively poor with respect to our previous report of 12 mW/$\pi$ \cite{lee2017demonstration}, and this is mainly due to the additional power consumption along the wires from the micro-heater to the bonding pad, which adds to the power dissipated at the contact resistance between the metal wire and the heater layers. We expect that the heating efficiency can be improved simply by increasing the length of the heater. In the circuit reported in Fig. \ref{fig:SiPIC-1}(c) the length of the waveguide in the MZI arms is 240 $\mu$m: this implies that photon-pairs can be generated also in the waveguide of the MZI of Fig. \ref{fig:SiPIC-1}(c). In order to isolate the waveguide contribution, we thermally tuned the microring out of resonance with respect to the pump photons (we label this situation RingOff). We changed the temperature of the microrings from (11.7$\pm$0.1)$^\circ$C (RingOn, microring resonant with the pump photons) to (20$\pm$0.1)$^\circ$C (RingOff) since the microring resonance wavelength temperature dependence of 80 pm/$^\circ$C \cite{lee2017demonstration}. The circuit shown in Fig. \ref{fig:SiPIC-1}(c) includes an additional second MZI: the first MZI, like in the circuits shown in Figs. \ref{fig:SiPIC-1}(a) and (b), contains the degenerate photon-pair sources and a phase shifter (${\phi_1}$), while the second MZI allows measuring the quantum interference of the generated photon pairs. Note that two multi-mode interference devices (1x2 MMIs) are used after the first MZI to tap and monitor the generated photons. The second MZI includes also a phase-shifter (${\phi_2}$). This is placed 740 $\mu$m away from the microrings to avoid any thermal cross-talk. At the output of the SiPIC the photons are out-coupled through gratings by the fiber arrays, filtered off the pump photons with band-pass filters (BPFs, with a 0.8 nm bandwidth centered at 1550.12 nm) and detected by using super-conducting-nanowire single-photon detectors (SNSPDs, EOS from Single Quantum) \cite{lee2022controlled-not, lee2019noise}. Then the SNSPD single-photon events are counted by a time-correlated single photon counter (TCSPC, Logic16 from UQdevices) and analyzed by logical post-selection.

Fig. \ref{fig:SiPIC-2}(a) shows schematically the experimental setup based on SiPIC-2. This is an improvement with respect to the circuit shown in Fig. \ref{fig:SiPIC-1}(c). It is aimed at a direct comparison between microring and waveguide sources by integrating both a HOM and a reverse HOM experimental set-up with a scheme similar to the one shown in Fig. \ref{fig:SiPIC-1}(c). In SiPIC-2, the microrings have 28.5 $\mu$m long radii, the spiral waveguides are 15 mm-long and the grating couplers have a measured coupling efficiency of 3.3 dB. In addition, SiPIC-2 includes photon pump filters realized by asymmetric MZIs (a-MZIs) and placed after the sources to reduce the accidental coincidence (noisy) counts. The asymmetric length of the a-MZI is designed to be 47.2 $\mu$m which yields a 12.8 nm (1600GHz) FSR to match four times the nominal FSR of the microrings.
Fig. \ref{fig:SiPIC-2}(c) shows the normalized transmission spectra measured with the photodiodes (PD$_1$-PD$_4$) visible in Fig. \ref{fig:SiPIC-2}(a). PD$_1$ and PD$_2$ allow measuring the pump filter characteristics, while PD$_3$ and PD$_4$ the transmitted photons. After tuning the resonance of the microring close to the pumping wavelengths (1543.78 and 1556.53 nm) and tuning the pump filter MZIs to reject the pump, the following data are observed: a loaded Q-factor of 3 x $10^4$ (twice larger than the Q-factor of the microrings in SiPIC-1), an FSR of 3.2 nm (400GHz, similar to the design value), a pump filter FSR of 12 nm (1500GHz), a pump rejection ratio larger than 30 dB. This last value is large enough to significantly reduce the accidental counts due to pump photons in coincidence measurements.

\section{Photon-pair generation and heralded single photons from microring resonators and waveguide spirals}
\label{sec:nonlin}

The overall quality of the integrated sources has been characterized by the measurements of the coincidence-to-accidental ratio (CAR) \cite{lee2019noise}, the heralding rate \cite{lee2019noise, lee2022controlled-not} and the $g^{(2)}(0)$ of the heralded photons or $g_h^{(2)}(0)$ \cite{Signorini2020on-chip}. To measure the $g_{h}^{(2)}(0)$ we added a 3dB fiber splitter between the SiPIC-1 outputs and the SNSPDs, realizing an Hanbury-Brown-Twiss interferometer \cite{Signorini2020on-chip}.

\begin{figure}[http]
    \centering
    a)\includegraphics[scale=0.27]{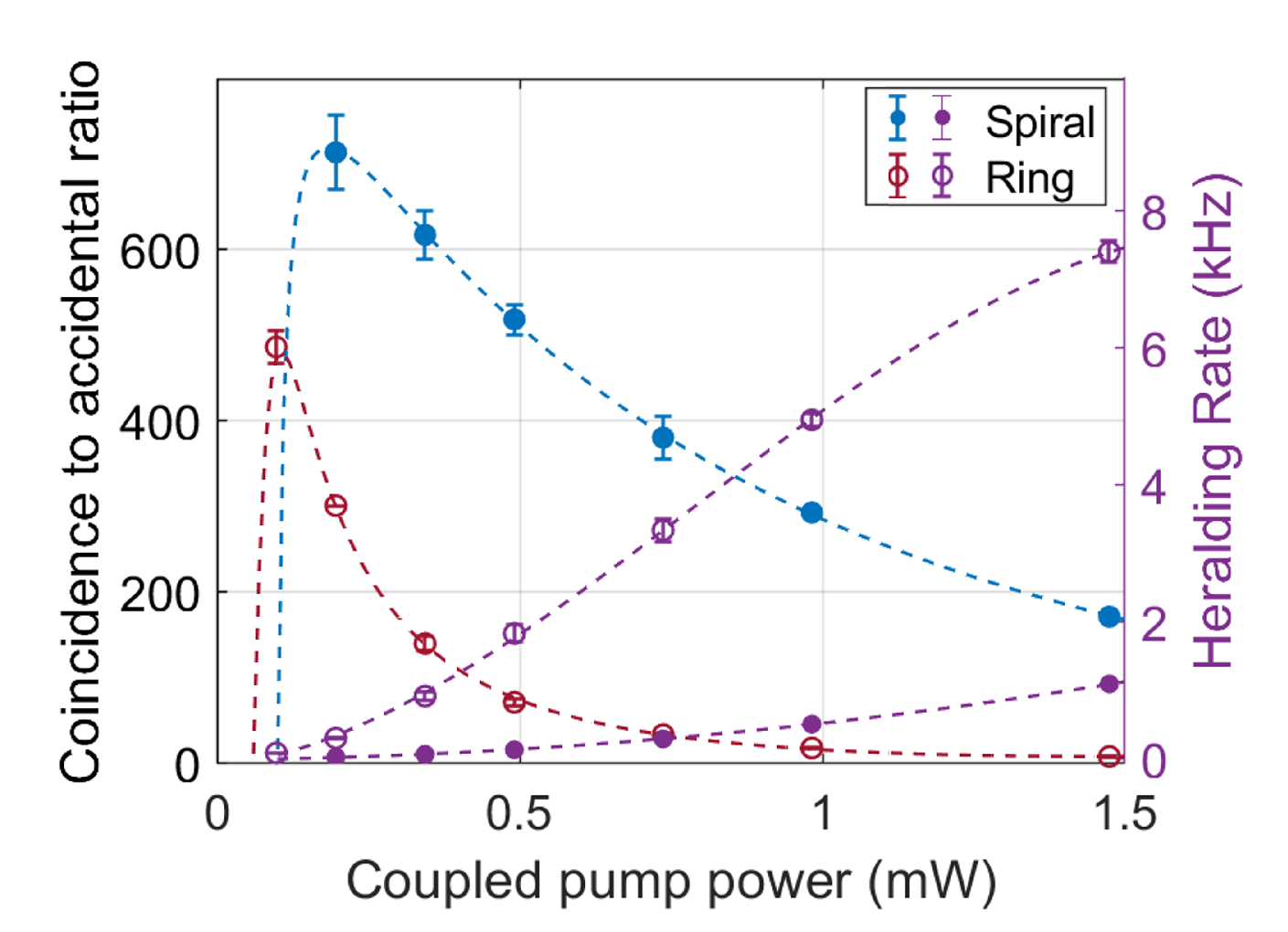} b)\includegraphics[scale=0.27]{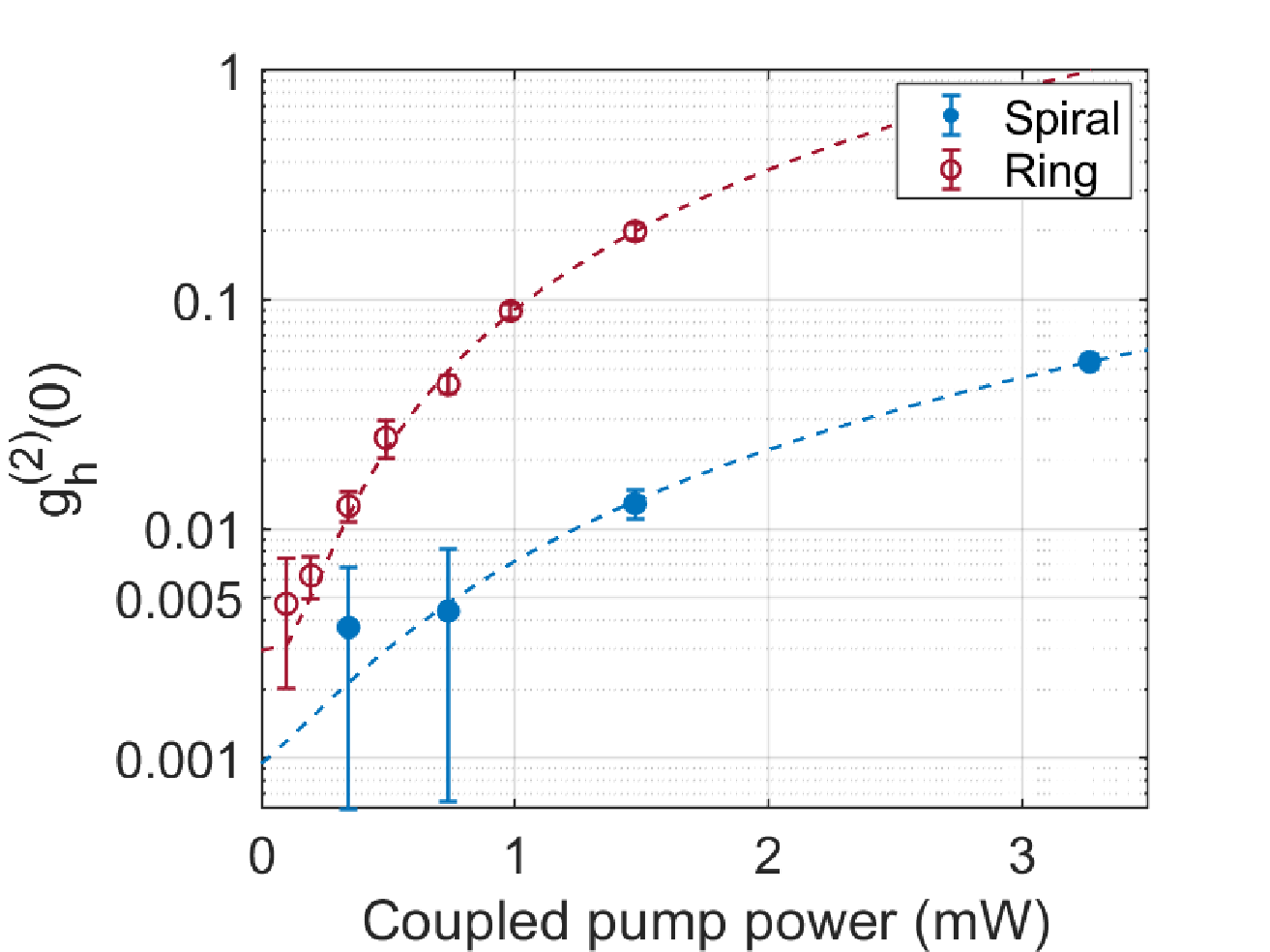}
\caption{\small{Measurements results of the characteristics of the photon-pair sources in the circuits shown in Fig. \ref{fig:SiPIC-1}(a) - spirals - and Fig. \ref{fig:SiPIC-1}(b) - rings. (a) Coincidence to accidental ratio as a function of the pump power coupled to the chip (blue dots refer to the spirals, empty red dots to the rings), and heralding rate as a function of the coupled pump power (violet dots refer to spirals, empty violet dots to microrings). (b) $g_h^{(2)}(0)$ as a function of the coupled pump power (blue dots refer to spirals, empty red dots to microrings).  } }
\label{fig:CAR_g2_firstcase}
\end{figure}

Fig. \ref{fig:CAR_g2_firstcase}(a) shows the measured CAR and heralding rate and Fig. \ref{fig:CAR_g2_firstcase}(b) the measured $g^{(2)}(0)$ of the heralded photons from the spiral waveguide sources (Fig. \ref{fig:SiPIC-1}(a), in short spirals in the following) and from the microring resonator sources (Fig.\ref{fig:SiPIC-1}(b), in short rings in the following) in SiPIC-1. 
The heralding rate for a microring resonator is 5 kHz for 1 mW pump power (-122 dB conversion efficiency for heralding), while the heralding rate for a spiral-waveguide is 0.5 kHz for 1 mW pump power (-129 dB conversion efficiency for heralding).
The coincidence time window of the TCSPC was set equal to 0.2 ns. Measurements show that the spirals have a larger CAR but a smaller heralding rate than the rings at high pump power. This last indicates that the field enhancement factor in our microring resonators is high enough to compensate for the length of the spirals. In addition, we observe in Fig. \ref{fig:CAR_g2_firstcase}(b) high-quality heralded single photons at low pump power (vanishing $g_{h}^{(2)}(0)$), while multi-photon contributions get relevant at high pump power (above 2 mW). The multi-photon contribution appears to be more severe for photon pairs generated from the rings than from the spirals.

\begin{figure}[http]
    \centering
    a)\includegraphics[scale=0.27]{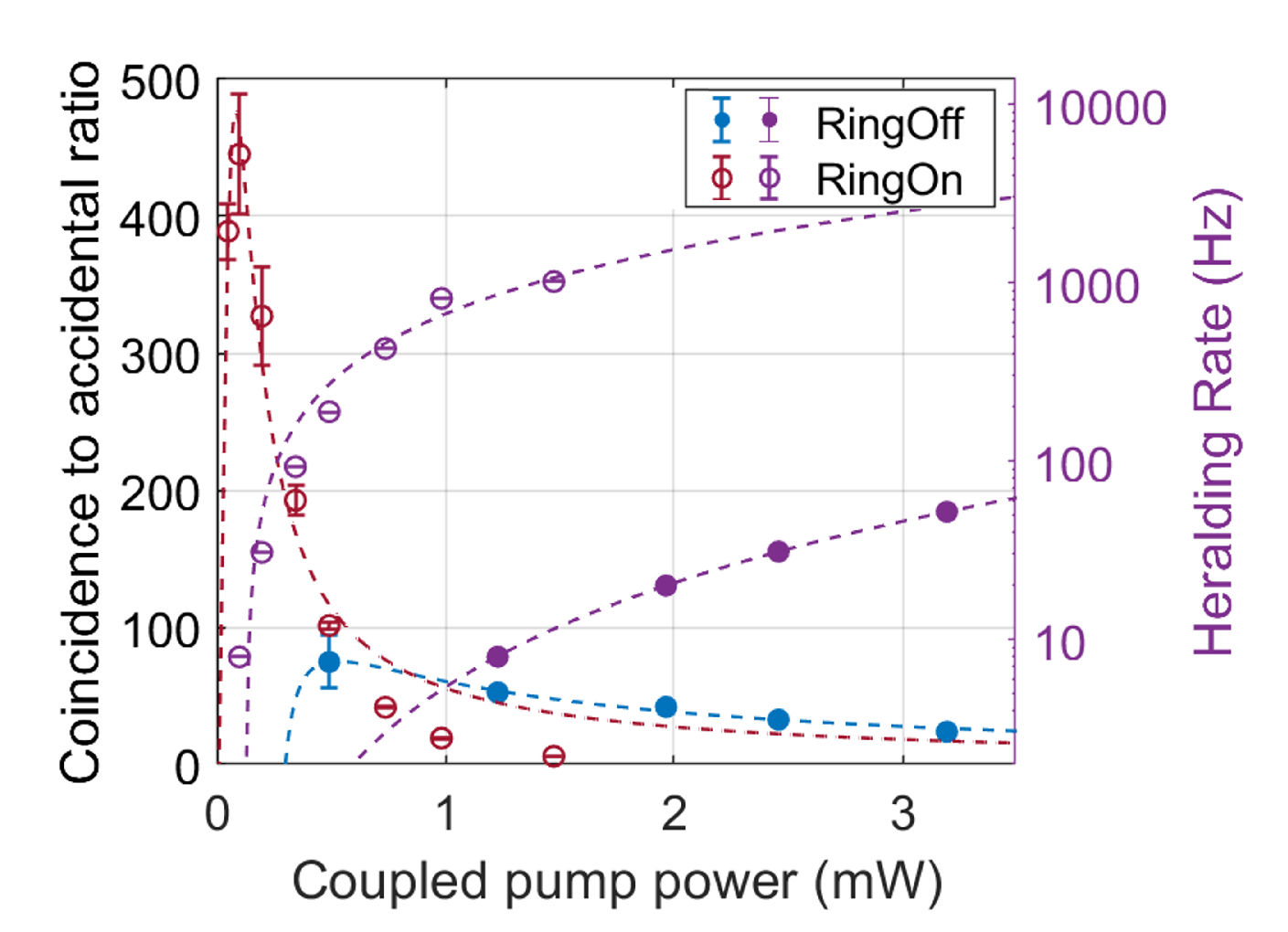}
    b)\includegraphics[scale=0.27]{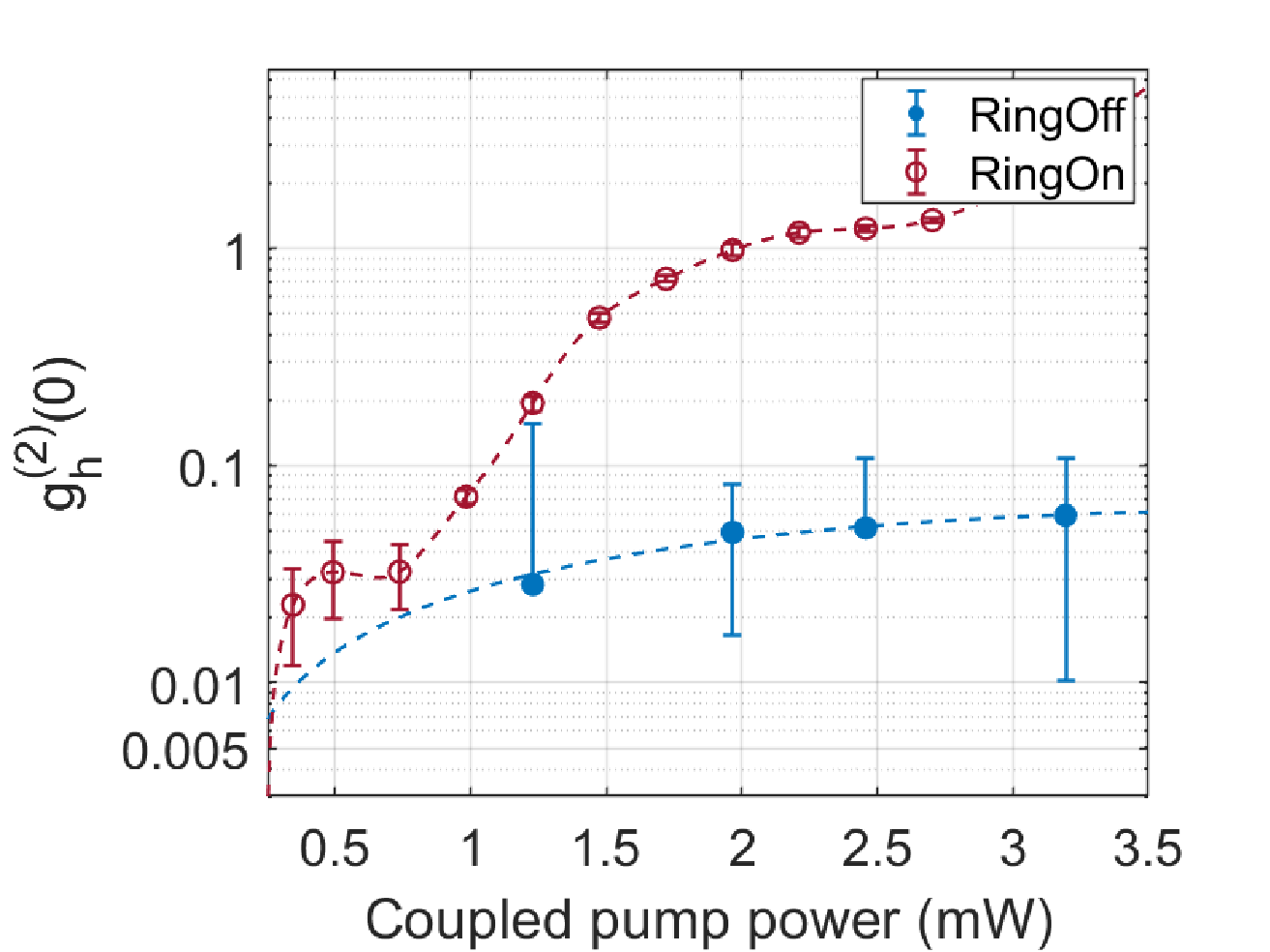}
\caption{\small{ Measurements of the characteristics of the photon-pair sources in the circuit shown in Fig. \ref{fig:SiPIC-1}(c). (a) Coincidence to accidental ratio as a function of the pump power coupled to the chip (blue dots refer to the RingOff mode - i.e. the microrings are off-resonant with the pump photons wavelengths, empty red dots to RingOn - the microrings are resonant with the pump photons wavelengths), and heralding rate as a function of the coupled pump power (violet dots refer to RingOff, empty violet dots to RingOn). (b) $g_h^{(2)}(0)$ as a function of the coupled pump power (blue dots refer to RingOff, empty red dots to RingOn). } }
    \label{fig:CAR_MZI_Spiral_and_Ring}
\end{figure}

The circuit in Fig. \ref{fig:SiPIC-1}(c) allows tuning the rings in resonance  (RingOn) and out of resonance (RingOff) with $\lambda_{P1}$ and $\lambda_{P2}$, and observing the photon-pair generations at the output channels ch1 and ch2. Fig. \ref{fig:CAR_MZI_Spiral_and_Ring} shows the CAR, the heralding rate and the $g_{h}^{(2)}(0)$ for the RingOn and RingOff configurations. It is observed in Fig. \ref{fig:CAR_MZI_Spiral_and_Ring}(a) that the heralding rate is larger for RingOn than for RingOff. The result for RingOn is similar to what is shown in Fig. \ref{fig:CAR_g2_firstcase}(a) for the ring, while the CAR and the heralding rate for RingOff are lower than those reported for the spirals in Fig. \ref{fig:CAR_g2_firstcase}(a) due to the difference in the waveguide lengths (240$\mu$m vs 15 mm), that enters quadratically in the photon-pair generation rate.
To measure $g_h^{(2)}(0)$, we set the two rings in the two arms of the first MZI in RingOn and RingOff modes, respectively, while we set $\phi_2=\pi$ of the second MZI. In this way, we can herald by ch2 detections the single photons at ch1 or ch3, i.e. by post-selection we measure the coincidences of ch1 and ch3 heralded by ch2 to get $g_h^{(2)}(0)$ (Fig. \ref{fig:CAR_MZI_Spiral_and_Ring}(b)). The results are similar to what can be observed in Fig. \ref{fig:CAR_g2_firstcase}(b) when the different waveguide lengths or microrings are considered.

\begin{figure}[http]
    \centering
    \includegraphics[scale=0.32]{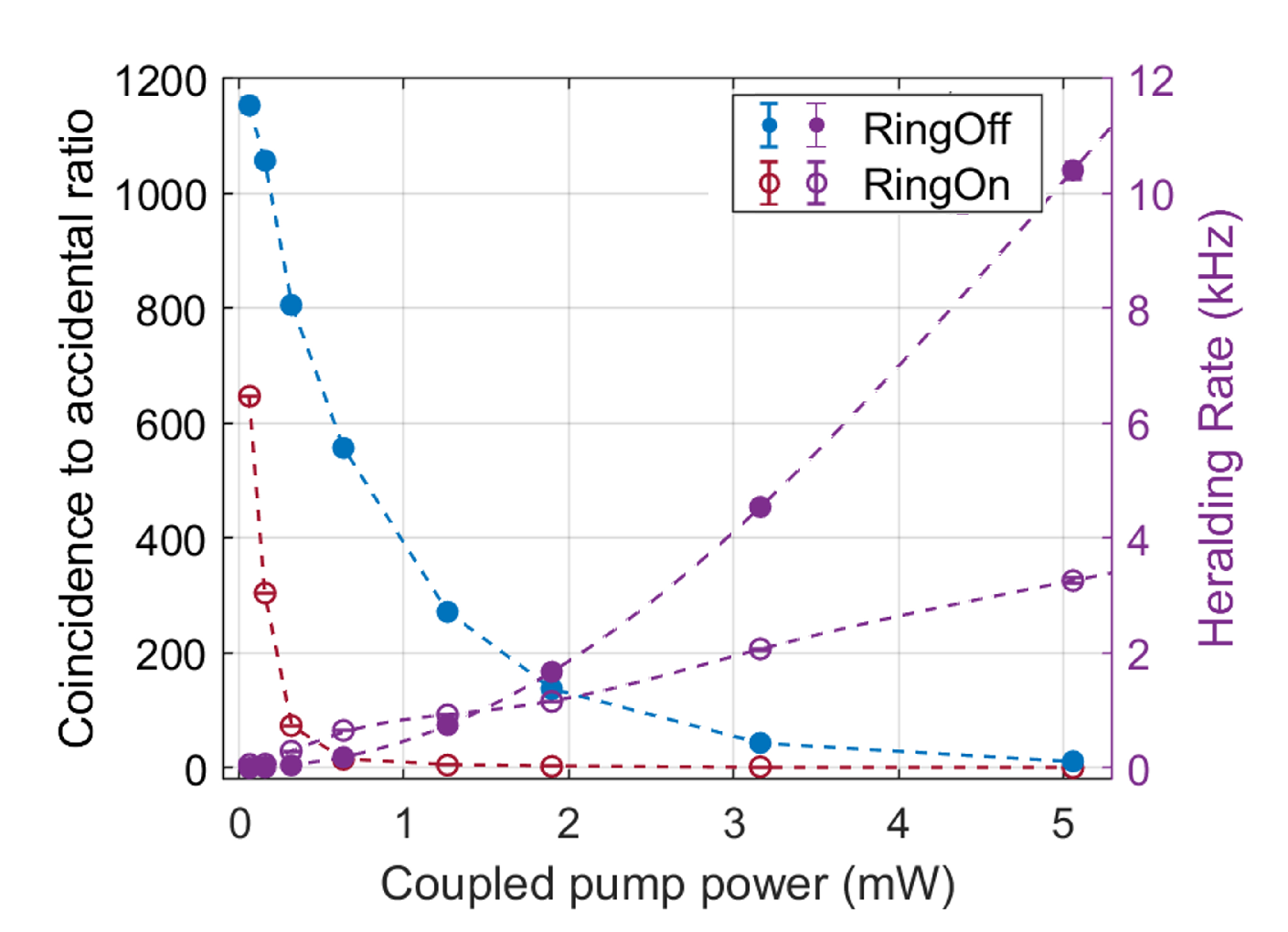}
\caption{\small{Measurements of the characteristics of the photon-pair sources in SiPIC-2: coincidence to accidental ratio as a function of the pump power coupled to the chip (blue dots refer to the RingOff mode - i.e. the microrings are off-resonant with the pump photons wavelengths, empty red dots to the RingOn mode - the microrings are resonant with the pump photons wavelengths), and heralding rate as a function of the coupled pump power (violet dots refer to RingOff, empty violet dots to RingOn).}}
    \label{fig:CAR_MZIringspiral}
\end{figure}

Data for SiPIC-2 are shown in Fig. \ref{fig:CAR_MZIringspiral}. During the measurements, the phase $\phi_2$ of the second MZI is fixed at zero (see Fig. \ref{fig:SiPIC-2}(a)) and, also in this case, the two modes RingOn and RingOff are possible. When the RingOn mode is selected the rings are effective as photon-pair sources, while in the RingOff mode the spirals are effective. The CAR is larger than 1000 for the spirals (600 for the rings) which demonstrates the effect of the removal of the pump photons by the on-chip pump filters based on the a-MZIs, as in our previous report \cite{lee2019noise}. However, the heralding rate for rings is relatively low due to the additional 3 dB losses caused by the 15 mm-long spirals which follow the microrings on the arms of the first MZI (a decrease of the coincidence probability by 1/2 x 1/2 = 1/4 is estimated), and due to large two-photon absorption (TPA) losses in the microring because of the high Q-factor.\\
Regarding TPA, we observe that Figs. \ref{fig:CAR_g2_firstcase}(a), \ref{fig:CAR_MZI_Spiral_and_Ring}(a) and \ref{fig:CAR_MZIringspiral} show a saturation of the quadratic behavior of the heralding rate with respect to the pump power. However, as it is shown in the following sections, our comparative experiments are run by using low pump powers to achieve optimal values of CAR and $g_h^{(2)}(0)$, where the quadratic behavior of the generation rate is observable. This sets us in a regime where TPA is not limiting the performances. Indeed, using low pump powers, the conversion efficiency of photon pair generation is low but nonlinear losses given by TPA and free carrier absorption can be neglected with respect to linear scattering loss \cite{Clemmen:09}. Thus, we can affirm that SFWM in SiPIC-1 and SiPIC-2, and especially for rings, should exclusively be performed at low optical pump powers to avoid a steep roll-off in performance at higher powers \cite{Ma:20}.

\section{
Reverse Hong-Ou-Mandel interference from waveguide spirals
}
\label{sec:pair}

\begin{figure}[http]
    \centering
a)\includegraphics[scale=0.28]{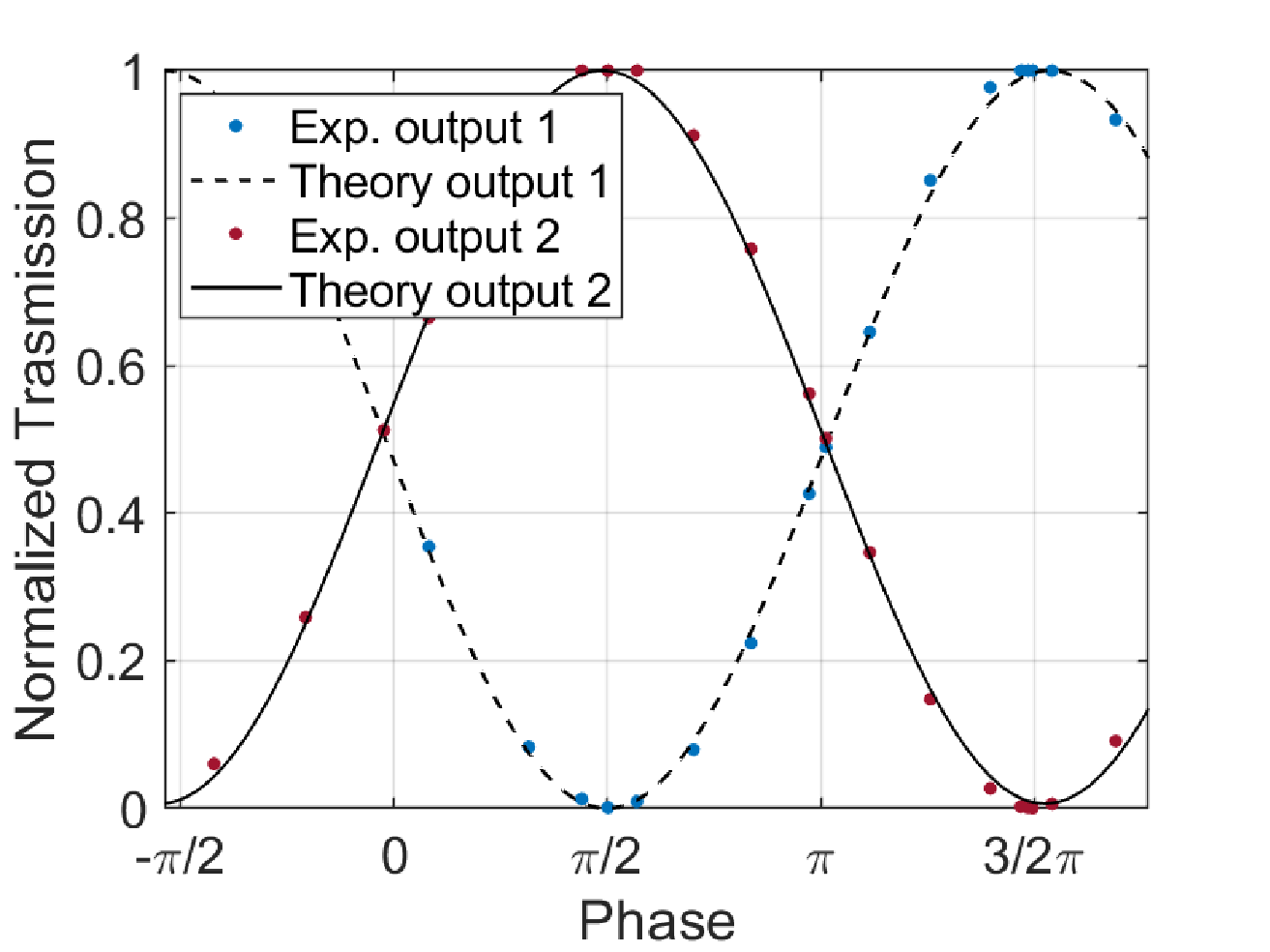} b)\includegraphics[scale=0.28]{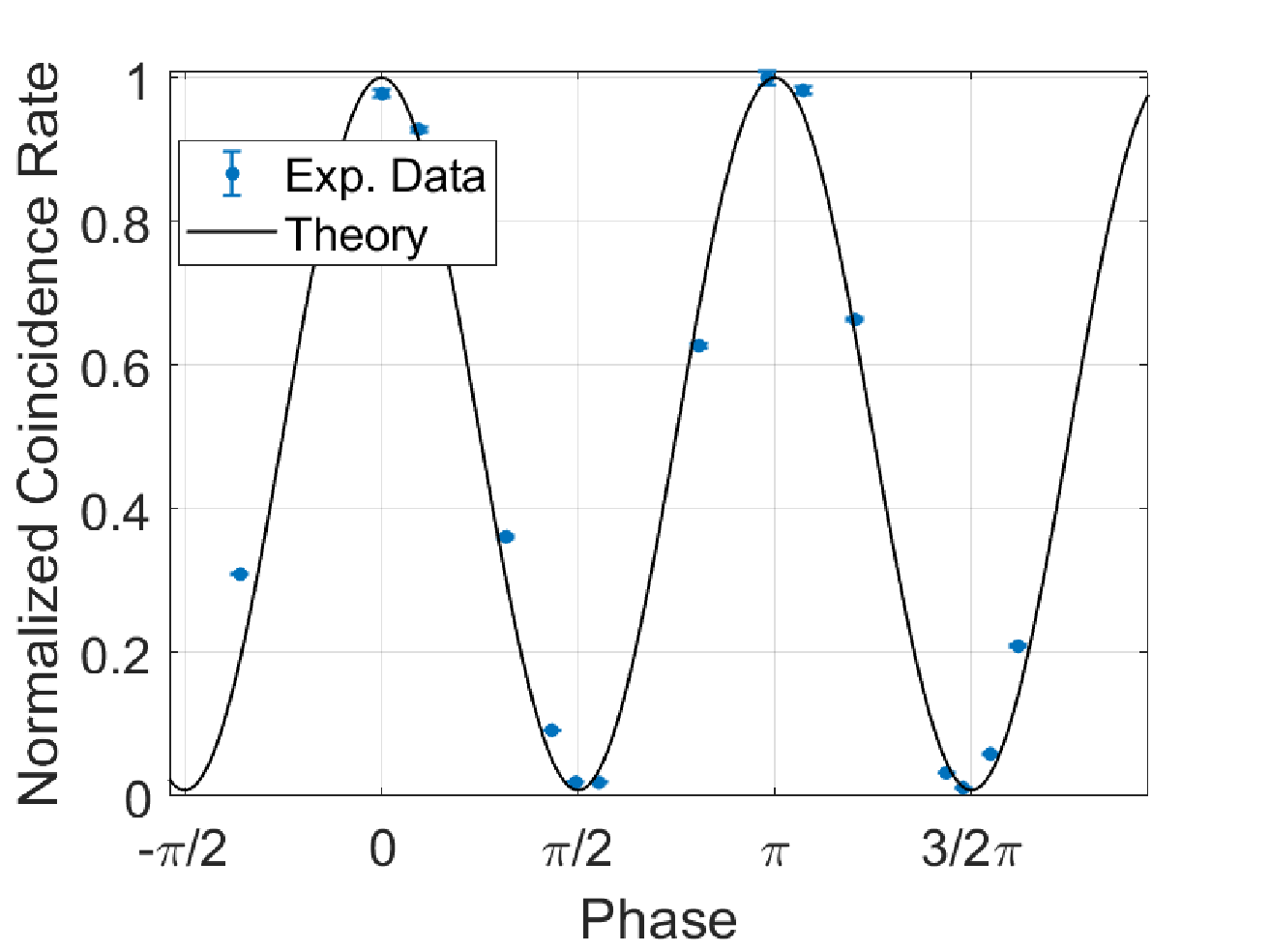}
\caption{\small{(a) Measured classical transmissions from the two outputs (blue dots ch1, red dots ch2) of the MZI in Fig. \ref{fig:SiPIC-1}(a) - SiPIC-1 - as a function of the phase $\phi$ compared to the theory (dashed line ch1, continuous line ch2). (b) Measured (blue dots) and theoretical (line) coincidence rates between the two outputs of the MZI  as a function of the phase ${\phi}$.}}
    \label{fig:revHOM_MZI_spiral}
\end{figure}

To prove the indistinguishability of the photons generated by the two sources, we performed quantum interference measurements with SiPIC-1 \cite{silverstone2014on-chip}. First, we studied the 15-mm long spiral-waveguide-based sources in the circuit of Fig. \ref{fig:SiPIC-1}(a) and we measured the dependence of the coincidence rates between ch1 and ch2 as a function of the phase $\phi$ of the MZI at a fixed pump power of 1.5 mW (CAR=170). Fig. \ref{fig:revHOM_MZI_spiral} compares the measured counts detected at the two output channels of the MZI in Fig. \ref{fig:revHOM_MZI_spiral}(a), showing the classical light transmission, with the coincidence rates between the two channels in Fig. \ref{fig:revHOM_MZI_spiral}(b), showing the quantum interference between the generated photons. As expected by the theoretical analysis reported in the Appendix or in \cite{silverstone2014on-chip, lee2022controlled-not}, the coincidence rates follow $\sin^2\left(\phi- \frac{\pi}{2}\right)=\cos^2(\phi)$, while the classical transmission follow $\sin^2\left(\left(\phi- \frac{\pi}{2}\right)/2\right)$ or $\cos^2\left(\left(\phi- \frac{\pi}{2}\right)/2\right)$. Some deviations with respect to the theory are observed due to the use of a first 1x2 MMI instead of a 2x2 MMI in the MZI. The coincidence measurement result is due to a reverse HOM quantum interference of degenerate photon-pairs at the second MMI of the MZI, and its high visibility of $98.8\%$ demonstrates the indistinguishability of the photon-pairs generated by the two spirals \cite{silverstone2014on-chip, Signorini2020on-chip}.
We use the following formula for the visibility ($V$): $V= \frac{R_c^{\rm max}-R_c^{\rm min}}{R_c^{\rm max}}$, where $R_c^{\rm max}$ and $R_c^{\rm min}$ are the maximum and minimum coincidence rates. In the Appendix, such formula is expressed in terms of coincidence probabilities in Eqs. \eqref{V_prob} and \eqref{V_prob_bis}.
Then, we used the rings in the circuit of Fig. \ref{fig:SiPIC-1}(b). However, because the micro heater of the phase shifter in the MZI is located too close (130 $\mu$m) to the microring, a large thermal cross-talk caused the microring resonance to shift out of resonance from the pump wavelength when the phase $\phi$ was varied. This impeded the measurement.
  
\section{On-chip direct comparison of microring resonators and waveguide spirals
}
\label{sec:ring}

The circuits of Fig. \ref{fig:SiPIC-1}(c) in SiPIC-1 and of \ref{fig:SiPIC-2}(a) in SiPIC-2 allow comparing the quantum interference of photon pairs produced by microring resonators or waveguides. Photons are generated in the first MZI while quantum interference measurements are performed by changing the phase $\phi_2$ of the second MZI. Actually, this corresponds to a complex HOM measurement sequence (a reverse HOM in the first MZI and a sequence of a HOM and a reverse HOM in the second MZI, see in particular Fig.\ref{fig:SiPIC-2}(a)).

\begin{figure}[http]
    \centering
    a)\includegraphics[scale=0.29]{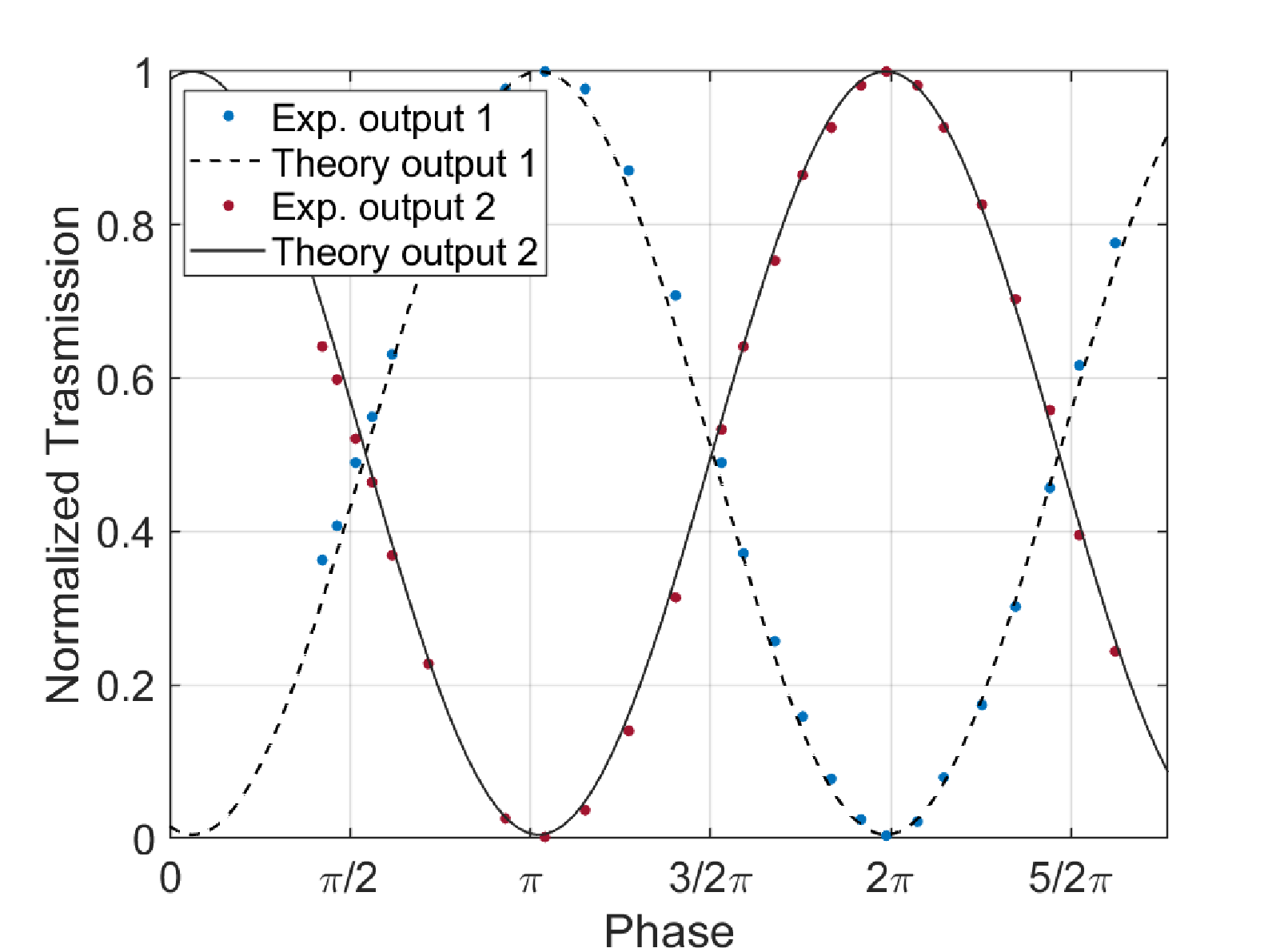} b)\includegraphics[scale=0.3]{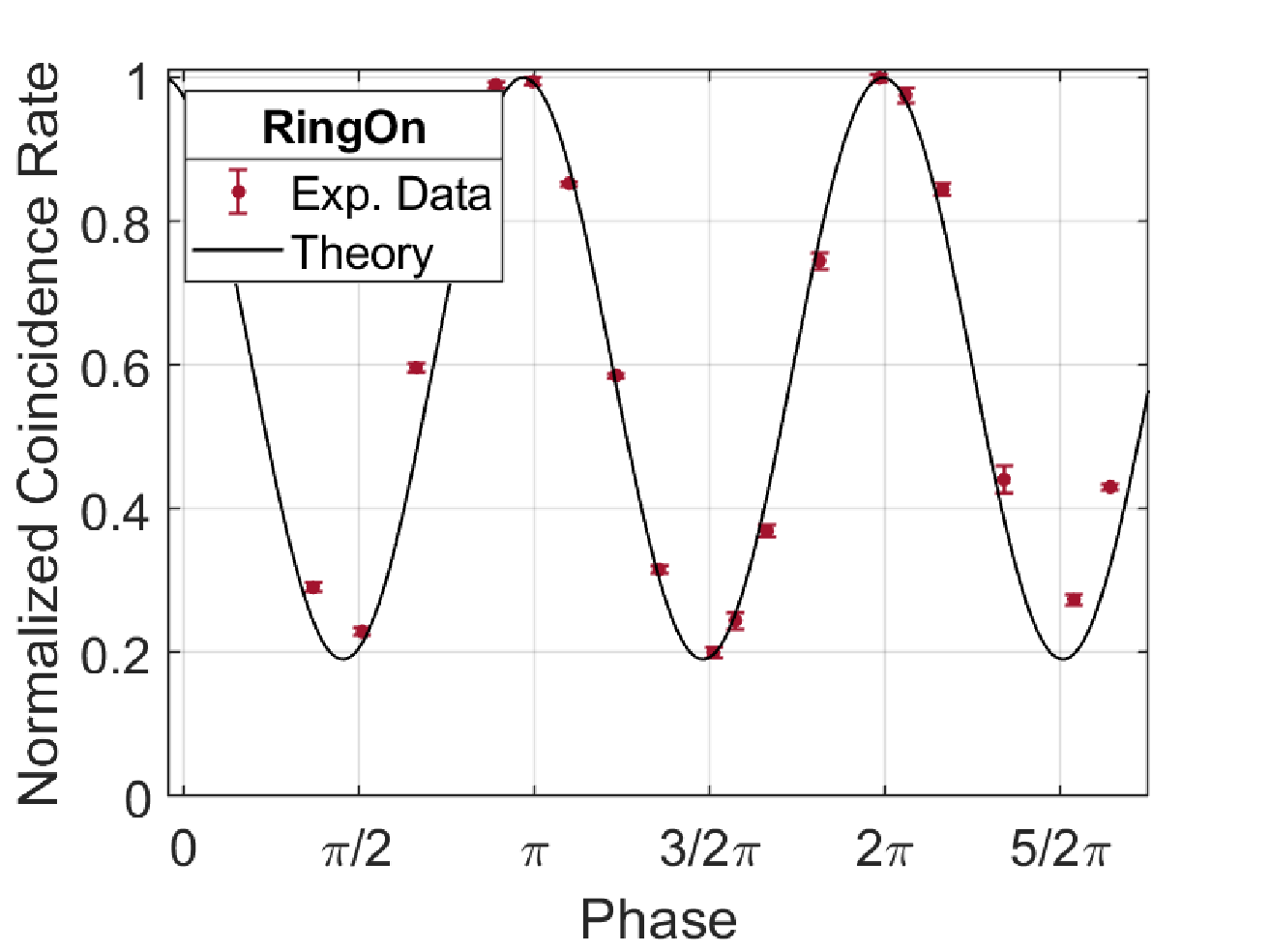}c)\includegraphics[scale=0.3]{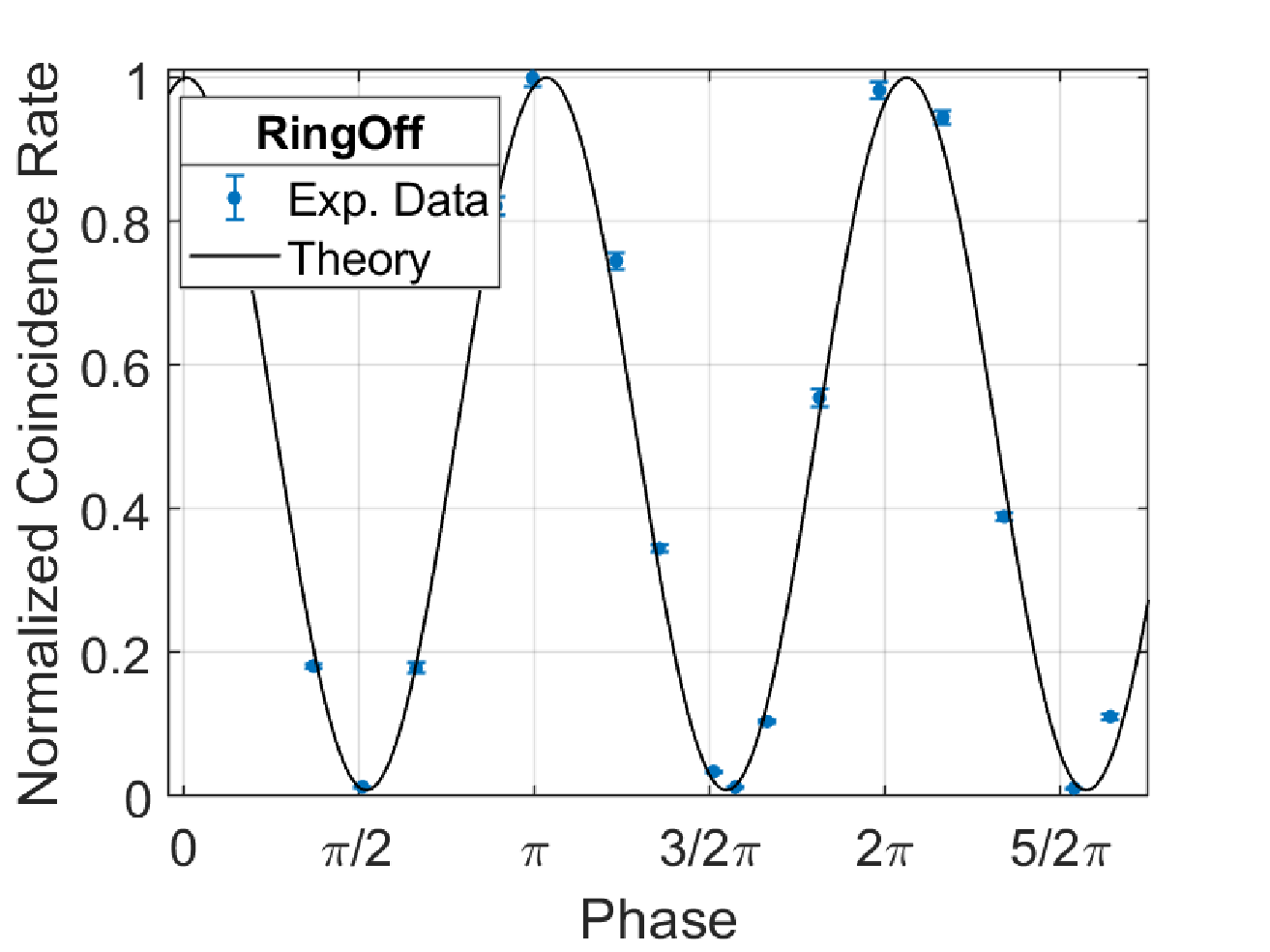}
\caption{\small{(a) Classical light transmissions at the outputs of the two channels of the second MZI in the circuit of Fig. \ref{fig:SiPIC-1}(c) - SiPIC-1 - as a function of the phase ${\phi_2}$ (dots measurements, lines theory). (b) Coincidence rates (red dots experiments, line theory) as a function of $\phi_2$ for the RingOn configuration. (c) Coincidence rates (blue dots experiments, line theory) as a function of $\phi_2$ for the RingOff configuration.}}
    \label{fig:HOM_Ring_and_Spiral}
\end{figure}

Let us first consider SiPIC-1. Given the CAR and $g_{h}^{(2)}(0)$ values measured for the two RingOn and RingOff configurations (Fig. \ref{fig:CAR_MZI_Spiral_and_Ring}), the pump power was set to 1 mW for RingOn (CAR=19, equivalent to 5\% of accidental counts) and to 3.2 mW for RingOff (CAR=24, equivalent to 4\% of accidental counts). Fig. \ref{fig:HOM_Ring_and_Spiral}(a) shows the measured classical-light transmissions from channel 1 (ch3 detector) and channel 2 (ch4 detector) of the second MZI as a function of $\phi_2$. Figs. \ref{fig:HOM_Ring_and_Spiral}(b) and \ref{fig:HOM_Ring_and_Spiral}(c) show the measured coincidence rates between channel 1 and channel 2 as a function of $\phi_2$ in the RingOn and RingOff configurations, which correspond to quantum interference of photons generated in rings and in spirals, respectively. Note that to compensate for slight variations in the photon-pair generation rate due to the thermal cross-talk (heat flow from the micro-heater on the phase shifter in the second MZI to the first MZI), we normalized the coincidence rates between ch3 and ch4 to the coincidence rates between ch1 and ch2 (see Fig. \ref{fig:SiPIC-1}(c)).
 
As expected and discussed in the Appendix, the classical-light transmissions behave as $\cos^2(\phi_2/2)$ or $\sin^2(\phi_2/2)$, while the coincidence rates behave as $\cos^2(\phi_2)$. For the RingOn configuration (Fig. \ref{fig:HOM_Ring_and_Spiral}(b)), a visibility of 80\% is observed which is lower than the 98.8\% visibility measured for the RingOff configuration (Fig. \ref{fig:HOM_Ring_and_Spiral}(c)).
The visibility for the RingOff is high considering the presence of 4\% of accidental coincidences at CAR=24. These additional coincidences are due to the classical interference of accidental photons coming from the residual pump, whose interference depends on the phase difference accumulated by the pump photons propagating from the first MZI to the second MZI. Therefore, considering a measured ratio of the ch3 and ch4 counts of about 3\% for $\phi_2 = \pi/2$ and 52\% for $\phi_2 = \pi$ implies that accidental coincidences are reduced at $\phi_2 = \pi/2$ yielding a high HOM visibility for the RingOff configuration as observed in Fig. \ref{fig:HOM_Ring_and_Spiral}(c). 

\begin{figure}[http]
    \centering
    a)\includegraphics[scale=0.27]{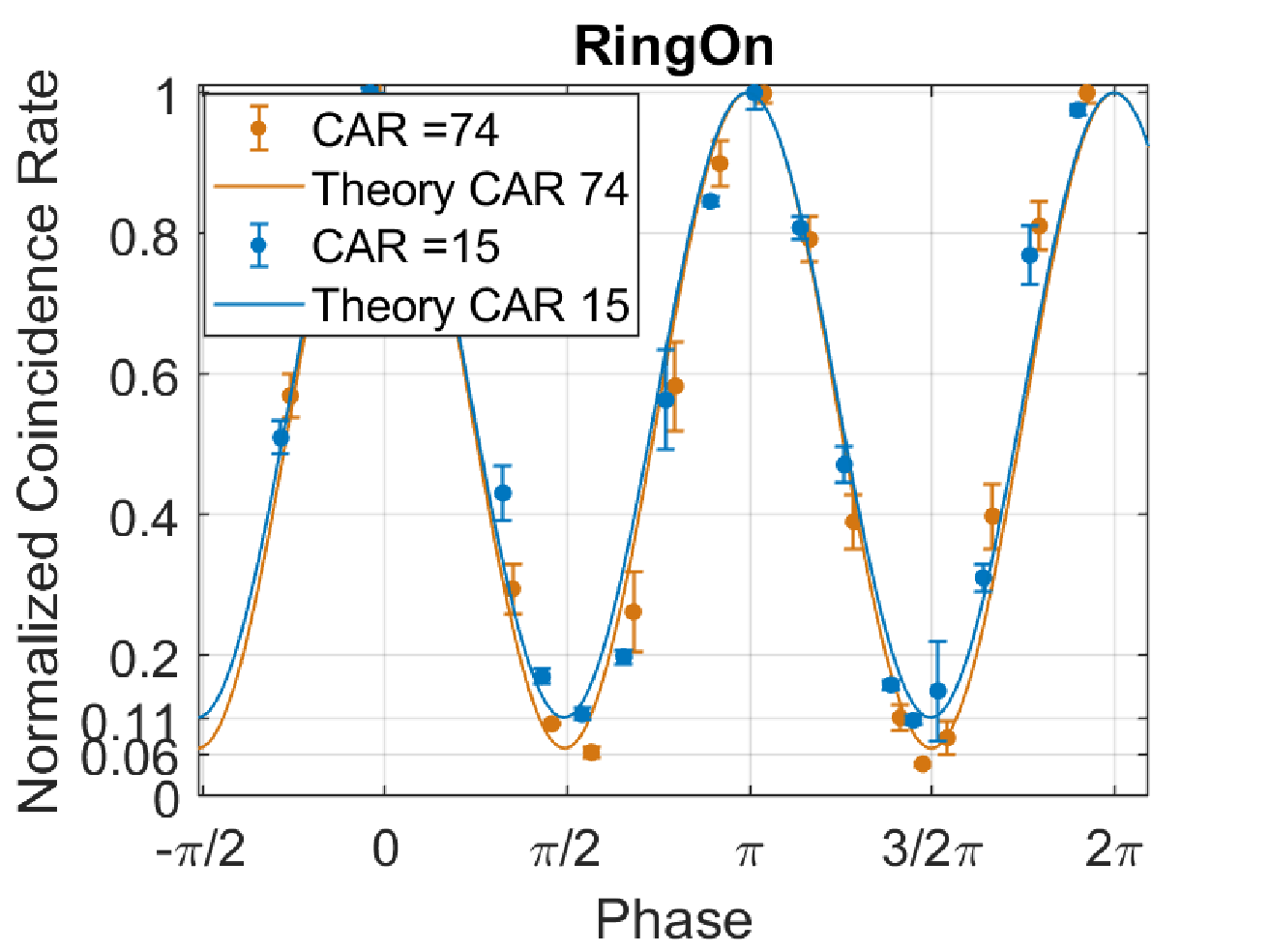}
    b)\includegraphics[scale=0.27]{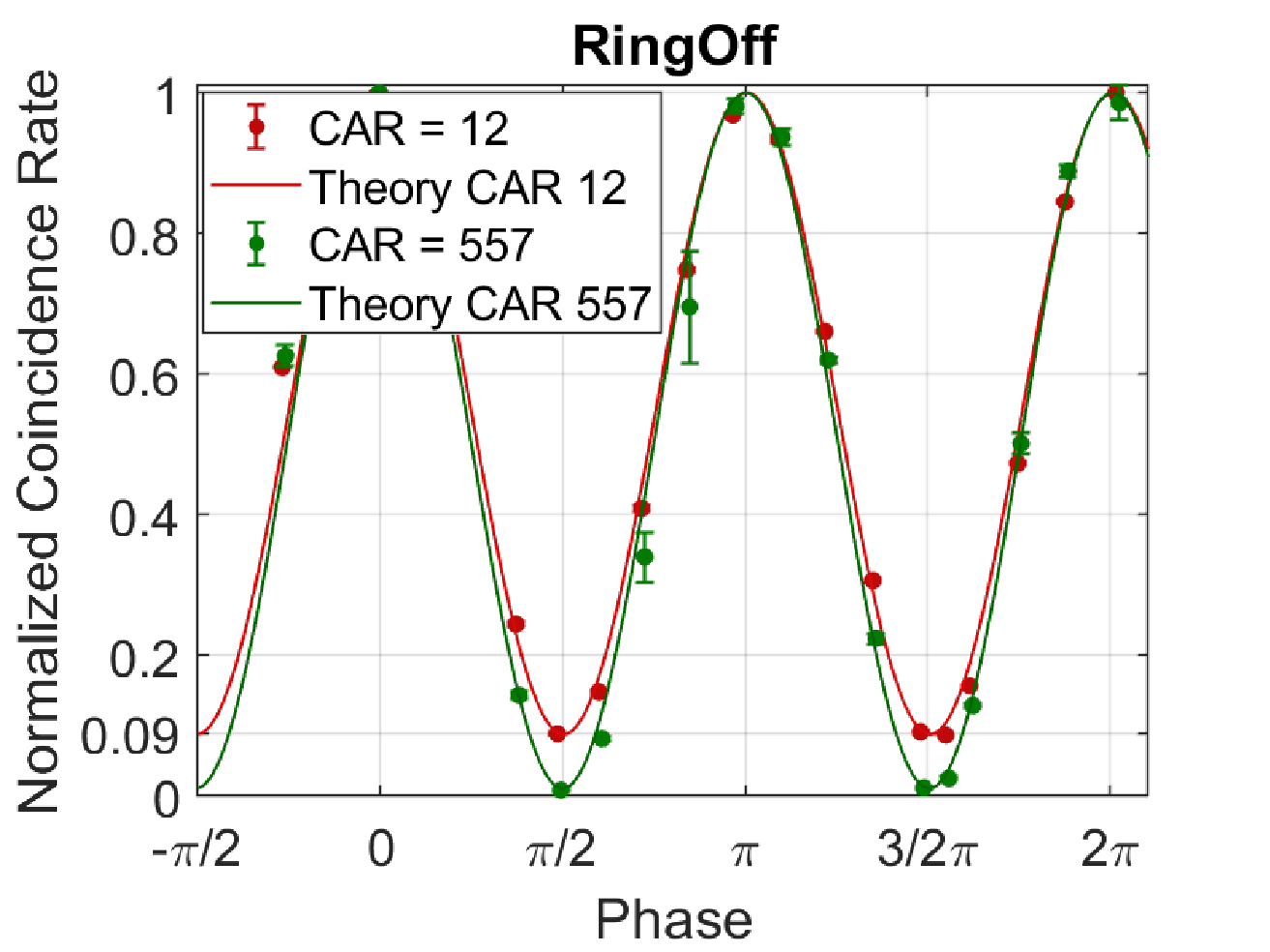}
\caption{\small{Coincidence rates between the output channels of the second MZI in the SiPIC-2 device (Fig. \ref{fig:SiPIC-2}(a)) as a function of phase ${\phi_2}$ of the second MZI. (a) RingOn configuration for two pump powers (orange 0.3 mW and blue 0.6 mW). Dots are experimental data, while lines are theoretical fits. (b) RingOff configuration for two pump powers (green 0.6 mW and red 5 mW). Dots are experimental data, while lines are theoretical fits.} }
    \label{fig:MZIringspiral_HOM}
\end{figure}

Fig. \ref{fig:MZIringspiral_HOM} shows the measured HOM interferences for the SiPIC-2 when the phase $\phi_2$ of the second MZI is varied (Fig. \ref{fig:SiPIC-2}(a)). 
For the RingOn configuration (Fig. \ref{fig:MZIringspiral_HOM}(a)), a HOM interference visibility of 94\% is measured when a pump power of 0.3 mW (corresponding to a CAR=74 in Fig. \ref{fig:CAR_MZIringspiral}) is used. The RingOn visibility decreases to 89\% for a pump power of 0.6 mW (CAR=15). Accidental counts impact at 1.3\% for CAR=74, and 6.3\% for CAR=15. So, the 94\% HOM interference visibility is equivalent to a 95.3\% visibility when the accidental noise is removed. For RingOff configuration (Fig. \ref{fig:MZIringspiral_HOM}(b)), HOM interference visibilities of 99\% for 0.6 mW pump power (CAR=557) and of 91\% for a pump power of 5 mW (CAR = 12) are observed. The accidental counts impact at 0.2\% for CAR = 557 and 7.7\% for CAR = 12.  So, the 99\% HOM interference visibility is equivalent to 99.2\% visibility when the accidental counts are excluded. Measurements show again that the HOM interference visibility depends on the CAR of the measured photon pairs. In addition, accidental photons due to residual pump classical interference yield a measured ch1 to ch2 ratio of 97\% for $\phi_2 = 0$ and of 90\% for $\phi_2 = \pi/2$ showing the effective role of the pump filters and a less relevance of this noise for SiPIC-2 than for SiPIC-1. Finally, the better HOM interference visibilities for SiPIC-2 than for SiPIC-1 are related to the enhanced CAR due to the integrated pump-filters and to an improved fabrication uniformity of the microrings in SiPIC-2 compared to SiPIC-1.

\section{Discussion}
\label{sec:disc}

The two types of photon-pair sources (waveguides and microring resonators) were simulated taking into account the input pump photon spectra as well as the band-pass filters (BPF) at the output channels. The joint spectral intensity (JSI) of the generated pairs was calculated as detailed in Appendix \ref{C}. Fig. \ref{JSI_fig} shows the JSI for the different sources in SiPIC-1 and SiPIC-2: 15-mm-long and 240-$\mu$m-long spiral waveguides, and microring resonators with Q-factors 1.5x$10^4$ and 3x$10^4$. In particular, it is worth noting that even in a degenerate SFWM process the generated photon pair wavelengths have a dispersion around $\lambda_s \simeq \lambda_i$ due to the spectral width of the pump laser lines and the generation band of the FWM process.

\begin{figure}[h!]
    \centering
    a)\includegraphics[scale=0.27]{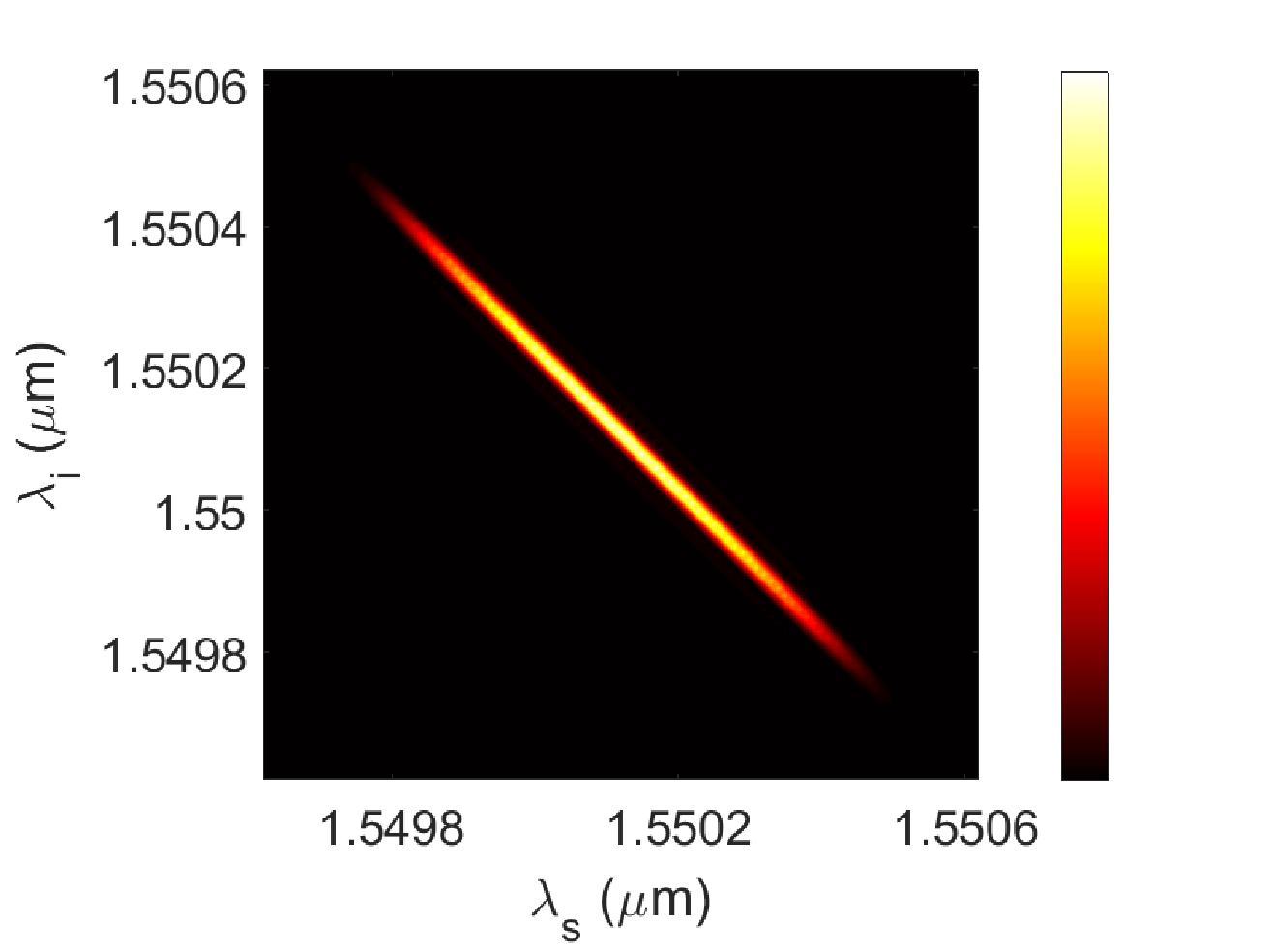}
    b)\includegraphics[scale=0.27]{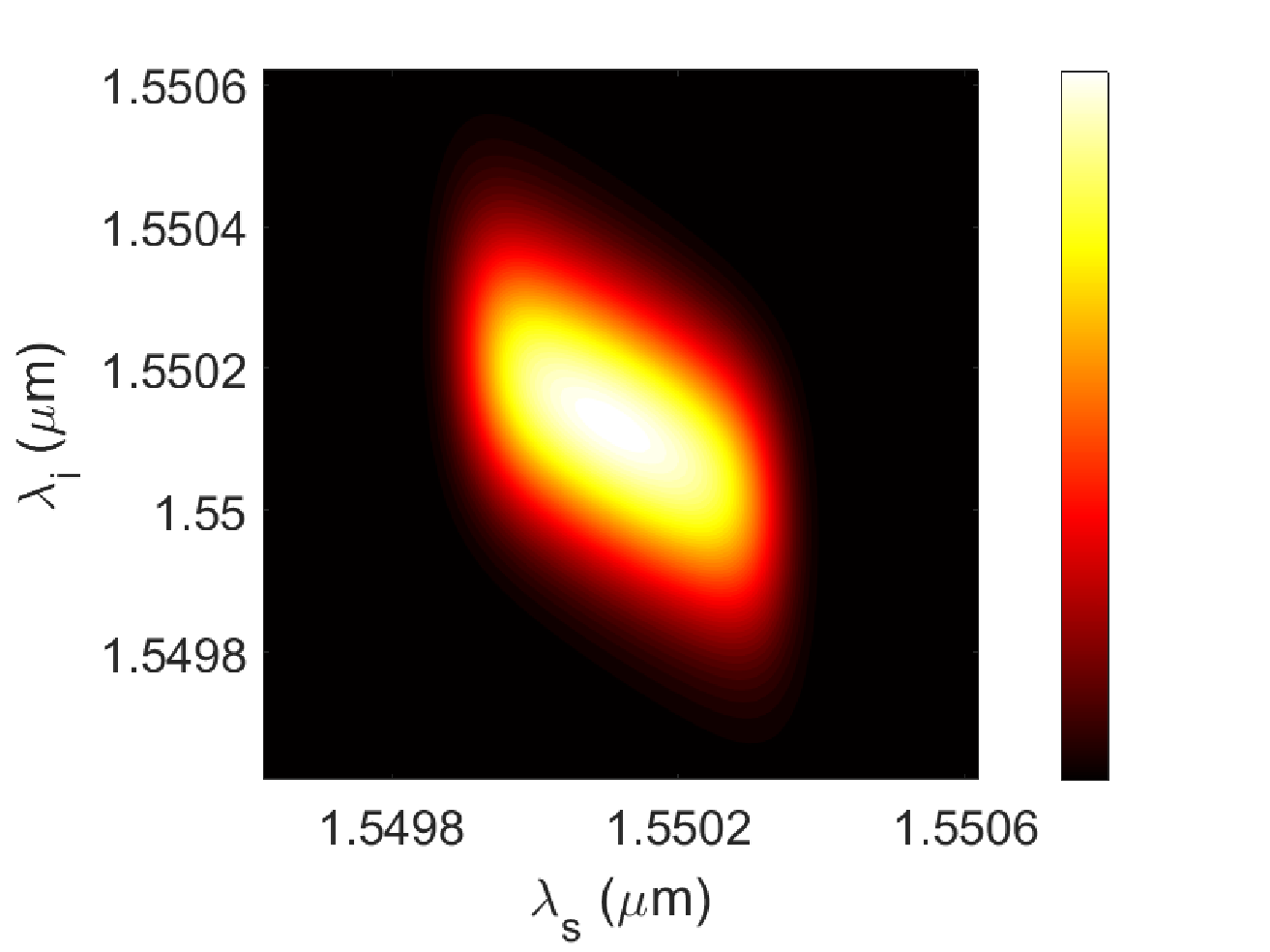}\\
    c)\includegraphics[scale=0.27]{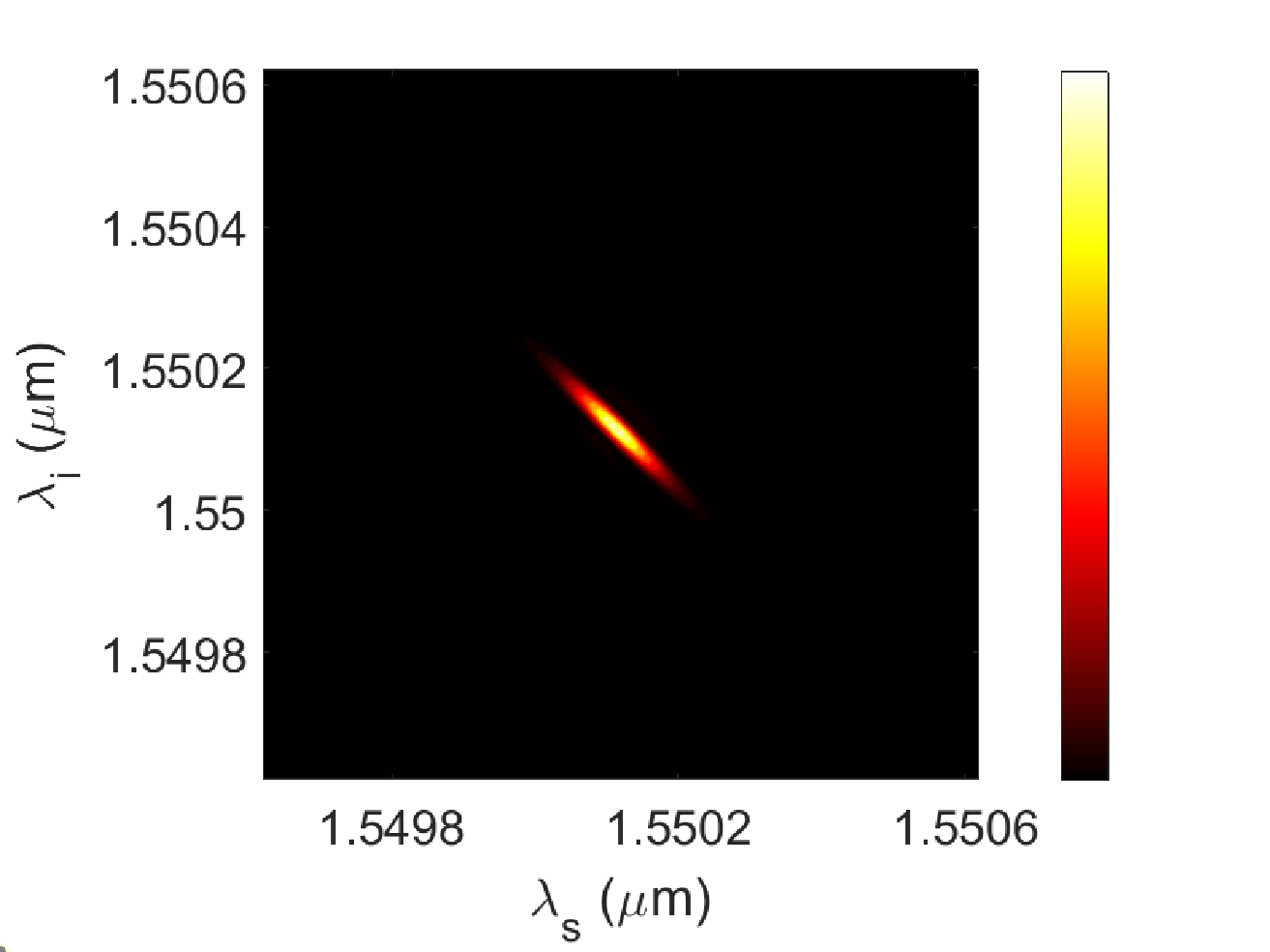} 
    d)\includegraphics[scale=0.27]{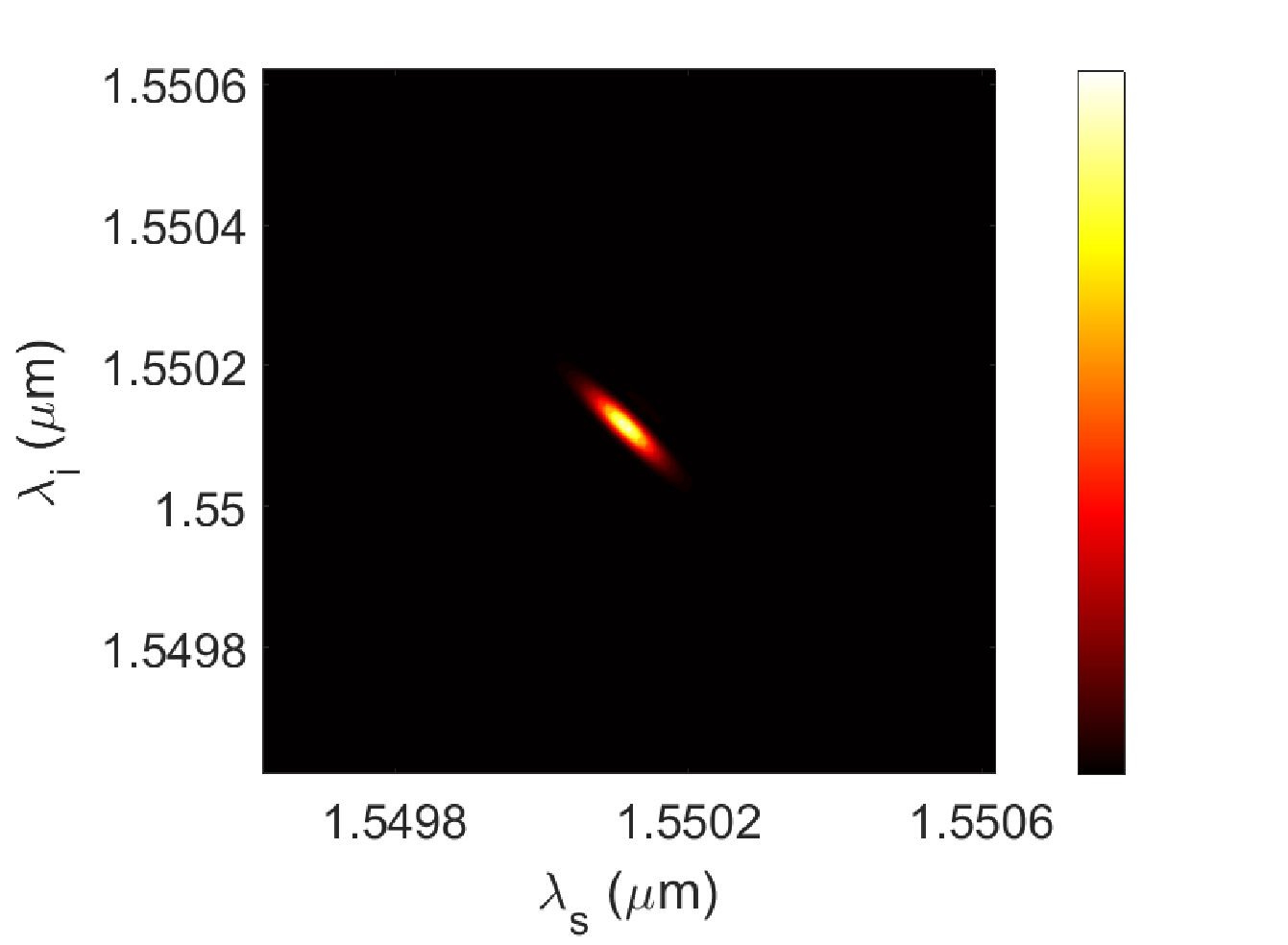} 

    \caption{Simulated JSI as a function of signal ($\lambda_s$) and idler ($\lambda_i$) wavelengths after the application of a band pass filter centered at $1550.12$ nm with a bandwidth of 100 GHz (0.8 nm) with CW pump lasers for: (a) 15-mm-long waveguide spiral, (b) 240-$\mu$m-long waveguide spiral, (c) microring resonator with an FSR around 3 nm and a Q-factor = 1.5x$10^4$ (0.1 nm FWHM), (d) microring resonator with an FSR around 3.2 nm and a Q-factor = 3x$10^4$ (0.05 nm FWHM). The simulations are based on parameters corresponding to the different circuits of Figs. \ref{fig:SiPIC-1} and \ref{fig:SiPIC-2}.}    
    \label{JSI_fig}
\end{figure}
The different shapes of the JSI for photon pairs generated by the spiral waveguides (Figs. \ref{JSI_fig}(a) and (b)) are mainly due to their different lengths. The longer waveguide acts as an additional narrow band-pass filter which modifies the JSI. This modification can be understood from the phase matching function reported in Eq. \eqref{phi} and contained in the joint spectral amplitude (JSA), Eq. \eqref{f}. It is also evident that the JSI shapes of the waveguides are wider than the one of the microring resonators. This implies a higher correlation in the generated photons from the waveguides than from the microring resonator. Thsese wide JSIs can be explained by the filtering of the broad waveguide generation band (between the two pump photon wavelengths) by the BPF. Filtering lowers also the photon-pair source brightness. In the case of the microring resonators instead, the effect of the BPF is negligible because of the narrow resonance spectral widths of the microrings. Moreover, their high Q-factor values make their JSIs narrow and increase their photon-pair generation probabilities, which depend by a cubic law on the Q-factor value \cite{helt2012does}. However, high Q-factor values make the microring resonator more sensitive to thermal crosstalk and TPA.

Table \ref{tab:summary} summarizes the measured and simulated properties of the two types of sources considered in this work. In the simulations for the JSI and purity, the two waveguides/microring resonators are assumed both identical to their nominal design values. This implies that the simulated values are not affected by the small variations of the nominal parameters present in the fabricated structures. The purity has been calculated by using Eq. \ref{Eq:purity} and the JSAs overlap has been calculated using the experimental results of the visibility, following the procedure discussed in the Appendix (\ref{app:RevHOM}-\ref{app:HOM-RevHOM}), and more specifically Eq. \eqref{eq:vis_vs_overlap}. 

In Appendix \ref{noise_intro}, we discuss how unwanted residual pump noise can be included into the description of the visibility in terms of the JSAs overlap. The qualitative effect consists in lower values of the visibility because of lower values of the signal-to-noise ratio. This behavior is found experimentally in Fig. \ref{fig:MZIringspiral_HOM}, where using an asymmetric MZI (aMZI) to filter the residual pump we observe different values of visibility for different CAR values. In Table \ref{tab:summary} we report the best values of observed visibility and the computation of the JSAs overlaps does not take into account the noise contribution discussed in Appendix \ref{noise_intro}. Thus, the reported value of computed JSAs overlap is a lower bound, since the assumption of negligible noise contribution might not be correct for our experiments.

\begin{table}[!h]
\begin{center}
\begin{tabular}{ |p{3.1cm}|p{2.9cm}|p{2.4cm}|p{3.2cm}|  }
  \hline
     & \centering Observed Visibility & \centering Simulated Purity &  Computed JSAs overlap\\
 \hline
15-mm waveguides (SiPIC-1) & \centering 98.8\% & \centering 81\% & \hspace{1.2cm}97.6\% \\
Microrings (SiPIC-1) & \centering 80\% & \centering 90\% & \hspace{1.2cm}66.6\% \\
0.24-mm waveguides (SiPIC-1) & \centering 98.8\% & \centering 86\% & \hspace{1.2cm}97.6\% \\
15-mm waveguides (SiPIC-2) & \centering 99\% & \centering 81\% & \hspace{1.3cm}98\% \\
Microrings (SiPIC-2) & \centering 94\% & \centering 90\%  & \hspace{1.2cm}88.7\% \\
 \hline
\end{tabular}
\caption{Summary of the relevant properties of the photon-pair sources in SiPIC-1 and SiPIC-2.}  
\label{tab:summary}
\end{center}
\end{table}

The purity in the case of the waveguide source goes from a simulated value of $20\%$ in the unfiltered case to the value reported in Table \ref{tab:summary} when a BPF is used. The simplicity in the fabrication of a waveguide source has its bottleneck in the necessity to use a high-quality BPF to produce photons with high spectral purity. Indeed, the results reported in the Table \ref{tab:summary} show similar values for the waveguide devices of SiPIC-1 and SiPIC-2 despite the fabrication of the latter resulted in devices with better performances. Indeed, the features of microring resonators greatly improved from SiPIC-1 to SiPIC-2 due to a better fabrication uniformity achieved with the second process. 

One can wonder why the visibility in the case of the waveguide is larger than the one of the microring resonator case, even if the purity shows an opposite trend. This point constitutes one of the main interesting aspects of the experiments described in Secs. \ref{sec:pair} and \ref{sec:ring}. In many published experiments the sources are independent \cite{farugue2018on-chip, llewellyn2020chip-to-chip,branczyk2017hong}, while in our case the two sources are not independent and therefore the visibility is not linked to the purity of the individual sources.
For example, in the case of independent sources, one typically uses non-degenerate SFWM to generate two pairs of idler and signal photons: the wave function is the product of the two pair wave functions and the heralding procedure is the partial trace with respect to the heralding twins. The visibility is then given by the purity of each heralded photon: for standard microring resonators this means that the visibility cannot exceed 92\% \cite{Helt:10,Vernon_2017}. 
Therefore, the property of being independent brings the overall wave function of the state to be separable, while in our case the state is not separable. To verify this, we can consider the state generated by the pair of sources, Eq. \eqref{eq:stateII} in the Appendix for small squeezing (the meaning of the different symbols is given in the Appendix),
\begin{align}
   |\Psi\rangle_{\rm II} 
    & \sim 
    \int \!\!
    {\rm d}\omega_i {\rm d}\omega_s
    \left[
    F_1(\omega_i, \omega_s)\,\,\hat a_1^\dagger(\omega_i)\hat a_1^\dagger(\omega_s)
    +F_2(\omega_i, \omega_s) \,\,\hat a_2^\dagger(\omega_i)\hat a_2^\dagger(\omega_s) \right]
   | {\rm vac} \rangle 
   \label{state_dep_sources}
\end{align}
where $F_{1}$ and $F_{2}$ are the JSAs of the two sources. Crucially, $|\Psi\rangle_{\rm II}$ is a superposition between pairs of photons generated in the two sources. 
What is generated in our case consists of a photon pair in a superposition of two paths, while in the case of independent sources and non-degenerate FWM two pairs are generated and after heralding a state with one heralded photon in each path is obtained.
In the Appendix, it is shown how the visibility is connected to the indistinguishability of the sources, represented by the JSAs overlap $\int \!\! {\rm d}\omega_i {\rm d}\omega_s    F_1(\omega_i, \omega_s) \bar F_2(\omega_i, \omega_s)$. In particular, given our definition of observed visibility, the visibility can be written as
\begin{align}
    V = \frac{2 \int \!\! {\rm d}\omega_i {\rm d}\omega_s    F_1(\omega_i, \omega_s) \bar F_2(\omega_i, \omega_s)}{1+\int \!\! {\rm d}\omega_i {\rm d}\omega_s    F_1(\omega_i, \omega_s) \bar F_2(\omega_i, \omega_s)}
    \label{eq:VandJSA}
\end{align}
which is presented in Eq. \eqref{eq:vis_vs_overlap} and derived in Appendix (\ref{app:RevHOM}-\ref{app:HOM-RevHOM}).
Therefore, even if the purity is low, what matters is how much the two sources are indistinguishable. This is quantified by the integral overlap between the two sources JSA and its relation with the visibility is given by Eq. \eqref{eq:vis_vs_overlap}.
We note that for $F_1=F_2$ the state generated by the couple of sources is exactly the input state to perform a reverse HOM experiment, see Eq. \eqref{state_dep_sources}, and Eq. \eqref{eq:VandJSA} gives $V=1$.
The result of visibility proves that there is a good matching between the spectral shapes of the photon-pair generated by the waveguides in SiPIC-1 and SiPIC-2 and by the microring resonators in SiPIC-2. 
The microring resonators in SiPIC-1 show a significantly lower JSAs overlap and a lower measured visibility: this can be due to the fact that some geometrical parameters can be slightly different between the two microring resonators. Therefore, the variability in the performances of the microring resonators between the two SiPIC gives an idea about the higher requirements on the fabrication process to reach a high level of uniformity in the components to assure proper indistinguishability of the different integrated sources when microrings are used. We note that the source purity is not affected by the Q-factor of the microring and reaches a theoretical maximum value of $92\%$ \cite{Helt:10,Vernon_2017}.

We conclude this section by reporting some numerical results obtained by simulating small deviations of one of the two sources with respect to the nominal design values.
In the case of the spiral-waveguide-based source, differences with respect to losses, length and width of the waveguides can be responsible for a lower JSAs overlap. These characteristics determine the generation band of the process since they enter into the phase matching function, Eq. \eqref{phi}. Let us note that the BPFs select $100$ GHz ($50$ GHz at 3dB bandwidth) around $1550.12$ nm, as it is described by the filters’ spectral amplitude in Eq. \eqref{eq:vis_vs_overlap}. The filtering acts as a quantum eraser \cite{quantum_eraser}, deleting the information about possible differences in the generation bands and making the structures less sensitive to deviations from nominal values. Thus, we expect that a narrower BPF will result in a higher visibility. Given the spectra of our filters, from our simulations we do not observe any significant change in the JSAs overlap by varying the relative width of the waveguides. Indeed, by considering a width variation of 5\% with respect to the nominal value, the JSAs overlap only decreases to 99.98\%. Instead, varying by 10\% the effective length $L_{\rm eff}$ of one of the two spiral waveguides with respect to the other, a JSAs overlap of 98\% is found. Comparing these values with Table \ref{tab:summary}, we infer that differences in propagation loss and/or length can be accounted as responsible for lower values of JSAs overlap. The case of microring-resonator-based sources is not affected by the action of the filters, and even if their footprint is much smaller, deviations in the coupling coefficients and ring losses can have a big impact on the JSAs overlap. In \cite{Oton:16}, it has been shown that devices uniformity within 1\% is achievable by using advanced processes in the same PIC. In a microring resonator, the coupling gap uniformity plays a very important role\cite{ Popovic:06, silicon_ring, Prinzen:13}, and it can determine a different coupling regime of the device. As an example, from our numerical simulations based on Eq. \eqref{fR}, we observe that the JSAs overlap decreases to 82\% for a 25\% Q-factor deviation, and to 67\% for a 50\% deviation. Fig. \ref{simu_imp} at the end of the Appendix shows the simulated JSAs overlap as a function of variations of $L_{\rm eff}$ for spiral waveguides and variations of Q-factor for microring resonators. Finally, the outcomes of our numerical analysis support our experimental results, where variations of the nominal design feature have a stronger effect on microring resonators by decreasing the JSAs overlap. Clearly, for both kinds of sources, large variations result in an increase of their distinguishability, allowing to answer the question ‘where did the pair generate?’ and suppressing the visibility of the HOM interference. Generally, we expect and observe that in the resonant case the impact of the non-uniformity of the devices is more severe.

\section{Conclusion}
\label{sec:conc}

We performed Hong-Ou-Mandel fringe interference experiments of degenerate photon pairs generated on-chip from nominally identical and non-independent probabilistic sources. Non-independency of the sources is achieved by putting them on the two arms of an MZI, while the HOM interference is realized by a second MZI. Through such a configuration, it is possible to simultaneously excite both sources and create a superposition of twin photons from the two sources, which is actually a path-entangled state. The photon-pair sources we studied and compared are microring resonators and waveguides. At best, we measured on-chip HOM visibility of 94$\%$ for microring resonators and $99\%$ for waveguides. The visibility of the HOM interference gives information about the indistinguishability of the twin photons generated, which is in turn the indistinguishability of the two nominally identical sources. 
Crucially, the non-independency of degenerate photon pairs sources results in the visibility being limited only by the indistinguishability and not by the purity. As a matter of fact, this places us in an ideal situation to address the question giving the title to this work, whose answer is the following: in quantum silicon photonics, spiral waveguides result to have better performances in indistinguishability than microring resonators. This result is quantified by the computed JSAs overlap of 98\% with waveguides and 89\% with microring resonators, see Table \ref{tab:summary}. 

The purity of the generated photon pairs plays a significant role when the generated photons have to interfere with photons produced by other independent photon sources \cite{farugue2018on-chip, llewellyn2020chip-to-chip}. However, in a generic quantum application it is desirable to have large values for both the purity and the indistinguishability because of the requirement of high visibility for dependent as well as independent sources. In our devices, we have seen two opposite cases: on one side a high visibility and JSAs overlap but a low purity and low brightness (waveguide-based photon-pair sources), while on the other side a low visibility and JSAs overlap but a high purity (microring based photon-pair sources). This prevents qualifying one source better than the other a priori. However, in the NISQ era of quantum computing one can simply choose what is needed to solve a task. Indeed, NISQ involves small numbers of qubits and algorithms that are focused on solving specific problems. 
One example is the Variational Quantum Eigensolver \cite{Peruzzo2014,Wang2018}, where what matters is the degree of indistinguishability and not the purity parameter. More generally, in applications where we create a photon pair in a superposition between different sources’ paths, similarly to what happens in our SiPICs, the quantum interference is influenced by how much indistinguishable the sources are.
On the other hand, in applications like Boson Sampling \cite{boson_sampling,boson_sampling_review}, where a large number of single (heralded) photons is required and the probability to inject a single photon state is proportional to the purity, one needs high values of purity.
This brings to the idea of {\it application-tailored sources}, i.e. sources designed for a specific algorithm and chosen to balance the required properties and the implementation efforts.
Following this idea, the ideal would be to have a library of photon-pair sources classified according to their main characteristics and choose the photon-pair source depending on the target class of problems and on the available fabrication resources.

From this point of view, our method can be understood as a quantifier for the indistinguishability of two sources and it can be generalized for the indistinguishability test of more than two sources, since the pair-wise test does not guarantee the overall indistinguishability of the sources \cite{Dufour_2020,PhysRevX.12.031033}. The generalization can be achieved by parallelizing more sources and bringing the superposition of pairs into a suitable network of MZIs. 
Related to that, other applications of the experimental scheme and the theoretical model can be found in the many-particle quantum interference investigations \cite{Dittel_2018,M_nzberg_2021,Dittel_2021}.

\begin{backmatter}
\bmsection{Funding}
ETRI (Grant No. 23YB1300), NRF funded by MSIT (Grant No. 2020M3E4A107845, 2022M3E4A1083526), Horizon 2020 project EPIQUS (Grant No. 899368).

\bmsection{Acknowledgments}
This work was supported by ETRI (Grant No. 23YB1300), NRF funded by MSIT (Grant No. 2020M3E4A107845, 2022M3E4A1083526), South Korea, EC within the Horizon 2020 project EPIQUS (Grant No. 899368) and by Q@TN, the joint lab between University of Trento, FBK- Fondazione Bruno Kessler, INFN- National Institute for Nuclear Physics and CNR- National Research Council.

\bmsection{Disclosures}
The authors declare no conflicts of interest.

\end{backmatter}

\appendix
\section{Appendix}
\setcounter{equation}{0}
\renewcommand{\theequation}{A\arabic{equation}}
\setcounter{figure}{0}
\renewcommand{\thefigure}{A\arabic{figure}}
\subsection{Joint spectral amplitude and single-mode squeezed state}
\label{C}

The joint spectral amplitude (JSA) describes how photon-pairs are correlated \cite{zielnicki2018joint, helt2012does}. It quantifies the probability density for the generation of one of the two photons in the $\omega_1$-frequency state given the second photon in the $\omega_2$-frequency state.
The generic bi-photon state takes the following form
\begin{align}
   |\Psi\rangle & = \int \!\! {\rm d}\omega_1{\rm d}\omega_2\,
   F(\omega_1,\omega_2)\hat a^\dagger(\omega_1) \hat a^\dagger(\omega_2) | {\rm vac} \rangle
   \label{eq:JSA}
   \\
   & \mbox{where}\,\, 
   \langle\Psi|\Psi\rangle = \int \!\! {\rm d}\omega_1{\rm d}\omega_2\, |F(\omega_1,\omega_2)|^2
   = 1
   \nonumber
\end{align}
$F$ is the JSA and $\hat a^\dagger(\omega)$ is the creation operator.
The joint spectral intensity (JSI) is the modulus square of the JSA, i.e. $|F(\omega_s,\omega_i)|^2$. JSI is directly linked to the intensity of the field, which is a measurable quantity.\\
Using the Schmidt decomposition for the JSA and two complete sets of orthonormal functions $\{u_{n}^{(1/2)}\}$ (the bar on a symbol stands for the complex conjugate of that quantity),
\begin{align}
   F(\omega_1,\omega_2)
   & = \sum_{\lambda} \sqrt{r_\lambda}\,u_\lambda^{(1)}(\omega_1) u_\lambda^{(2)}(\omega_2) 
   \label{eq:Schmidt}
   \\
   & \mbox{where}\,\, \int \!\! {\rm d}\omega \, u_{\lambda_1}^{(1/2)}(\omega)\,
   \bar u_{\lambda_2}^{(1/2)}(\omega)
   = \delta_{\lambda_1 \lambda_2}
   \,\,\,\mbox{and}\,\, \sum_\lambda r_\lambda
   = 1 \,,
  \nonumber
\end{align}
the bi-photon state can be written as a sum of product states
\begin{align}
   |\Psi\rangle = \sum_{\lambda} \sqrt{r_\lambda}\,
   \left(\int \!\! {\rm d}\omega_1\,u_\lambda^{(1)}(\omega_1) \hat a^\dagger(\omega_1)\right)
   \left(\int \!\! {\rm d}\omega_2\,u_\lambda^{(2)}(\omega_2) \,
    \hat a^\dagger(\omega_2) \right)| {\rm vac} \rangle \,.
    \label{state_Schmidt}
\end{align}
The set of coefficients $\{r_\lambda\}$ are called the Schmidt coefficients. Finally, the purity of the bi-photon state is given by
\begin{align}\label{Eq:purity}
    P = \int \!\! {\rm d}\omega_1{\rm d}\omega_2{\rm d}\omega_1'{\rm d}\omega_2' \,
    F(\omega_1,\omega_2) \bar F(\omega_1,\omega_2') 
    F(\omega_1',\omega_2') \bar F(\omega_1',\omega_2)
    =\sum_\lambda r_\lambda^2 \,.
\end{align}
This parameter is less or equal to 1 and quantifies how much the state is factorizable. 
Note that taking the partial trace of the density matrix corresponding to the state in Eq. \eqref{state_Schmidt}, we obtain 
\begin{align}
   \hat \rho_2 &:= \mbox{Tr}_1 |\Psi\rangle \langle\Psi| 
   = \int \!\! {\rm d}\omega_2{\rm d}\omega_2'
   \left( \int \!\! {\rm d}\omega_1\,
   F(\omega_1,\omega_2)\bar F(\omega_1,\omega_2')
   \right)
   \hat a^\dagger(\omega_2)  | {\rm vac} \rangle
   \langle {\rm vac}|
   \hat a(\omega_2') \\
   &= \sum_{\lambda} r_\lambda\,
   \left(\int \!\! {\rm d}\omega_2\,u_\lambda^{(2)}(\omega_2) \,
    \hat a^\dagger(\omega_2) \right)| {\rm vac} \rangle
    \langle {\rm vac}| \left(\int \!\! {\rm d}\omega_2'\,\bar u_\lambda^{(2)}(\omega_2') \,
    \hat a(\omega_2') \right)\,
    . \nonumber
\end{align}
From the previous equations, it is manifest that we have a pure state after the application of a partial trace on the bi-photon state when only one Schmidt coefficient is equal to one and all the others are null.
Therefore, when we are dealing with a source of bi-photon states, the purity parameter gives a quantitative information about how much the state of a single photon in the pair is pure.
It is an intrinsic property of the state and can be extended to the source. In particular, in the process of heralding it is used to express the quality of the source of photons in the production of single photon states.

Another important characteristic of photon sources is their indistinguishability, which is an extrinsic property since it is evaluated through the comparison of different sources. A good estimator for it is given by the overlap between the JSAs of the bi-photon states produced by different sources: this idea can be naively associated with the comparison of two normalized vectors through their scalar product.

It is natural to use the previous concepts in nonlinear spontaneous parametric processes, where the bi-photon state is generated by converting pump photons into the correlated signal and idler pair of photons.
In particular, the 
SFWM process is described by the following Hamiltonian \cite{helt2012does,Quesada:22}
\begin{align}
H_{\rm FWM} &=
- \frac{\gamma_{S I P_1 P_2} \,\hbar^2 \,\omega_{S I P_1 P_2} }{4\pi^2}
\int\!\! {\rm d} x \, {\rm d}\omega_1 \, {\rm d}\omega_2 \, {\rm d}\omega_3 \, {\rm d}\omega_4 \,
{\rm e}^{{\rm i} (\omega_1+\omega_2-\omega_3-\omega_4)t}
\\
& \hspace{2cm}
{\rm e}^{-{\rm i} (k_S(\omega_1)+k_{I}(\omega_2)-k_{P_1}(\omega_3)-k_{P_2}(\omega_4))x}
a^\dagger_{s}(\omega_1) a^\dagger_{i}(\omega_2) a_{P_1}(\omega_3) a_{P_2}(\omega_4)
+\mbox{h.c.}
\nonumber
\end{align}
where 
\begin{align}
\omega_{S I P_1 P_2} &= \left( \omega_{s} \omega_{i} \omega_{P_1} \omega_{P_2} \right)^{1/4} \,\,\,,
\hspace{0.8cm}
\gamma_{S I P_1 P_2} = n_2 \,
\frac{\omega_{S I P_1 P_2}}{c\,\mathcal{A}^{\rm eff}_{S I P_1 P_2}} \,,
\end{align}
 $\mathcal{A}^{\rm eff}$ is the effective area of the process,  $k$ is the wavevector of the four photons, $n_2$ is the nonlinear refractive index of the material and $\gamma$ is the parameter that characterizes the strength of the FWM process \cite{Quesada:22}.

In SFWM, we can have two different cases: the non-degenerate case where the pump photons are degenerate in frequency and the generated pair is not, and the degenerate case where we have the opposite situation.
In the experiment, we are interested in the case where two pump photons ($P_{1/2}$) are converted into two photons ($s$ and $i$) with the same frequency. This means that our JSA is a symmetric function and $u := u^{(1)} = u^{(2)}$.

In the low gain regime and for coherent laser pumps (thus the pump is treated classically), the final state is a squeezed state,
\begin{align}
    |\Psi \rangle = \hat{\mathcal{U}}\,
   | {\rm vac} \rangle \,,
   \quad \mbox{where}\quad
   \hat{\mathcal{U}} = \exp\left\{ \frac{1}{2} \xi \int \!\! {\rm d}\omega_s{\rm d}\omega_i\,
   F(\omega_s,\omega_i)\hat a^\dagger(\omega_s) \hat a^\dagger(\omega_i) -{\rm h.c.}\right\}
   \label{intro_state_fwm}
\end{align}


where $\xi$ is the squeezing parameter, $F(\omega_s,\omega_i)$ is the bi-photon wavefunction or JSA. In our case, the JSA can be written as
\begin{align}
  F(\omega_s, \omega_i) = \int \!\! {\rm d}\omega \, \alpha(\omega)\beta(\omega_s+\omega_i-\omega)\phi(\omega_s,\omega_i,\omega),
   \label{f}
\end{align}
where $\alpha(\omega)$ and $\beta(\omega)$
are the complex amplitudes of the pump beams, the function $\phi$ is the phase matching function, given by the following relation
\begin{align}
 \phi(\omega_s,\omega_i,\omega) &= \exp \left(\frac{i\Delta k(\omega_s,\omega_i,\omega) L}{2}\right) \,{\rm sinc}\left(\frac{\Delta k(\omega_s,\omega_i,\omega) L}{2}\right)\,,
   \label{phi} 
\end{align}
$L$ is the waveguide length and $\Delta k$ is the phase mismatch parameter \cite{garay2007photon}. 
As we can see from \eqref{f}, the JSA receives two contributions: one from the shape of the pump beams and one from the phase matching function that contains the kinematic parameter $\Delta k$.
In the microring resonator case \cite{Helt:10,Christensen_2018}, Eq. \eqref{f} takes a slightly different form to account for the enhancement effect due to the microring, in particular
\begin{align}
    F(\omega_s, \omega_i) = \textit{l}_s(\omega_s)\textit{l}_i(\omega_i) \int \!\! {\rm d}\omega \, \alpha(\omega)\textit{l}_{p_1}(\omega)\beta(\omega_s+\omega_i-\omega)\textit{l}_{p_2}(\omega_s+\omega_i-\omega) ,
    \label{fR}
\end{align}
where $\textit{l}_j(\omega)$ is the Lorentzian function relative to the microring resonance linewidth of the $j$-th resonance involved in the process.
Again, the JSA quantifies the probability density for the creation of the pair's first photon in the $\omega_s$ mode and the second photon in the $\omega_i$ mode. \\
Using \eqref{eq:Schmidt}, we can write the state as
\begin{align}
    |\Psi \rangle &= \exp\left\{ \frac{1}{2} \sum_\lambda \xi_\lambda \hat A_\lambda^2 -{\rm h.c.}\right\} | {\rm vac} \rangle
    =
    \bigotimes_\lambda \left[ 
    \frac{1}{\sqrt{\cosh \xi_\lambda}}
    \sum_{n=0}^\infty \left( \tanh \xi_\lambda\right)^n \frac{\sqrt{(2n)!}}{2^n n!} | 2n \rangle_\lambda
    \right] \, ,\\
    \mbox{where} & \quad \xi_\lambda := \xi \sqrt{r_\lambda} 
    \quad,\quad 
    | n \rangle_\lambda := \frac{1}{\sqrt{n!}}\left( \hat A_\lambda \right)^n |{\rm vac}\rangle
    \quad \mbox{and} \quad     
    \hat A_\lambda:=\int \!\! {\rm d}\omega\,
    u_{\lambda}(\omega) \,\hat a^\dagger(\omega) \,. 
\end{align}

The squeezing parameter contains the information about the pair-generation probability: note that by expanding \eqref{intro_state_fwm} for small squeezing and keeping only the first non-trivial term we have the same form of \eqref{eq:JSA}.

From the following statistic distribution, it is possible to obtain the main properties
\begin{align}
    \langle \hat n \rangle &=
    \sum_\lambda \left(\sinh \xi_\lambda\right)^2
    \\
    p_{\rm trig} &= 1-
    \sum_\lambda {\rm sech} \,\xi_\lambda
\end{align}
where $p_{\rm trig} $ is the probability to activate a threshold detector. Finally, we are now ready to take into account the effect of the losses.
Each lossy component is modeled as an ideal component preceded by an ideal beamsplitter with a non-unit transmission probability,
\begin{align}
     \hat A_\lambda \to \eta_\lambda \,  \hat A_\lambda + \sqrt{1-\eta_\lambda^2} \, \hat e_\lambda
\end{align}
where $0 \le \eta_\lambda \le 1$ being the transmission and $\hat e_\lambda$ is the destruction operator of the environment.
The full state, after accounting for losses and tracing over environment modes, can therefore be written as
\begin{align}
    \hat\rho &= \otimes_\lambda \,\hat \rho_\lambda \,,
    \\
    \hat \rho_\lambda &= 
    \frac{1}{\cosh \xi_\lambda}
    \sum_{n=0}^\infty  \left( \tanh \xi_\lambda\right)^{2n} \left(\frac{(2n)!}{2^{n} n!} \right)^2
    \sum_{k=0}^n 
    \frac{\eta_\lambda^{2(2n-k)} (1-\eta_\lambda^2)^{k}}{k!(2n-k)!}
    | 2n-k \rangle_\lambda
    \langle 2n- k |_\lambda \,.
    \nonumber
\end{align}
The previous properties are then modified in the following way
\begin{align}
    \langle \hat n \rangle &=
    \sum_\lambda \eta_\lambda^2 \left(\sinh \xi_\lambda\right)^2
    \,,
    \\
    p_{\rm trig} &= 1-
    \sum_\lambda \frac{{\rm sech} \,\xi_\lambda}{\sqrt{1-(1-\eta_\lambda)^2 \left(\tanh \xi_\lambda\right)^2}} \,.
\end{align}

\subsection{Evolution state description in the reverse HOM interference experiment}
\label{app:RevHOM}

In this section, we examine the theoretical underpinnings of the reverse HOM interference phenomenon \cite{hong1987measurement, branczyk2017hong, bouchard2020two} as seen in the integrated system shown in Fig. \ref{fig:SiPIC-1}(a). Therefore for simplicity, we restrict ourself to the waveguide source case only. The extension to the microring source is straightforward. 
In this subsection, photon creation through the SFWM process is described using the notation contained in the previous subsection. 
\begin{figure}[http]
    \centering
    \includegraphics[scale=0.4]{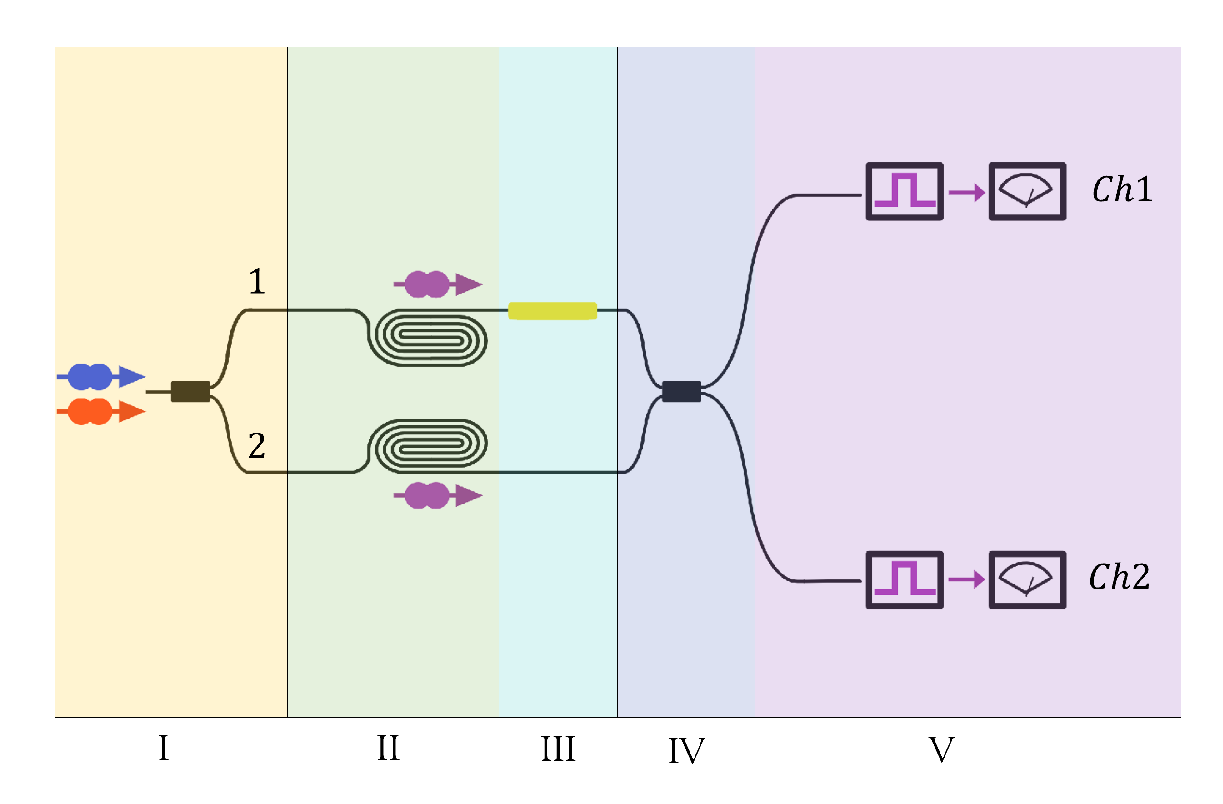}

\caption{\small{The scheme presented in Fig. \ref{fig:SiPIC-1}(a) is shown. Its division into five parts is beneficial for its analytical description.}
}
    \label{design_small}
\end{figure}
In order to follow the evolution of the state step by step, we divide the integrated silicon chip into five distinct sections, denoted by the Roman numerals \textbf{I} through \textbf{V} as shown in Fig. \ref{design_small}. In Sec. \textbf{I}, the pump photons are divided into two paths, path $1$ (upper) and $2$ (lower), by the first 1X2 MMI. In Sec. \textbf{II}, the degenerate SFWM process produces the pair of photons along the two paths. Through the use of a phase shifter, path $1$ in Sec. \textbf{III} gains a relative phase with respect to path $2$. A 2X2 MMI in Sec. \textbf{IV} causes interference between the photons propagating in the two arms. Finally, the filtering and detection of the two pathways are carried out in the final Sec. \textbf{V}.
\\
The initial state composed of two pump beams at the two frequencies, $\omega_{P1}$ and $\omega_{P2}$, can be described as:
\begin{align}
   |\Psi\rangle_0 = 
   \exp\left[\int \!\! {\rm d}\omega
   \left( \alpha(\omega) +\beta(\omega) \right) \hat A^\dagger(\omega) \right] | {\rm vac} \rangle
\end{align}
where $\hat A^\dagger(\omega)$ is the creation operator, $\alpha$ and $\beta$ are the pumps amplitude functions centered at $\omega_{P1}$ and $\omega_{P2}$. 
Note that we are working with unnormalized states, therefore any meaningful quantity must result from a proper ratio.

Because of the filter at the entrance of the chip, we can assume that
\begin{align}
    \int \!\! {\rm d}\omega \,\,\alpha(\omega) \bar\beta(\omega) = 0 \,.
\end{align}

In \textbf{Sec. I}  the  pump photons interfere with a 1X2 MMI. 
By applying this evolution to the initial state, we obtain:
\begin{align}
   |\Psi\rangle_{\rm I} = 
   \exp\left[\int \!\! {\rm d}\omega
   \frac{\left( \alpha(\omega) +\beta(\omega) \right)  }{\sqrt{2}} \left(\hat a_1^\dagger(\omega) + \hat a_2^\dagger(\omega) \right) \right] | {\rm vac} \rangle
\end{align}
where $\hat a_1^\dagger$ corresponds to the upper waveguide and $\hat a_2^\dagger$ to the lower one.
\\
In \textbf{Sec. II} the FWM mechanism generates correlated photons that annihilate two pump photons.
As we explained in the previous subsection, the state is 
\begin{align}
   |\Psi\rangle_{\rm II} 
    & = 
    \exp\left\{ \frac{\xi}{2} \int \!\!
    {\rm d}\omega_i {\rm d}\omega_s
    \left[
    F_1(\omega_i, \omega_s)\,\,\hat a_1^\dagger(\omega_i)\hat a_1^\dagger(\omega_s)
    +F_2(\omega_i, \omega_s) \,\,\hat a_2^\dagger(\omega_i)\hat a_2^\dagger(\omega_s) \right]
    \right\}| {\rm vac} \rangle 
    \label{eq:stateII}
\end{align}
where $F_{1/2}$ is the JSA of the source in the upper/lower waveguides. We neglect the residual contribution from the pumps and the non-degenerate generation process since our detection will capture only the contributions around the degenerate generation process.

In \textbf{ Sec. III}, a phase shifter is introduced into path $1$. This enables the photons from that path to undergo an additional phase shift. The resultant state will be:
\begin{align}
    |\Psi\rangle_{\rm III}
    = & 
    \exp\left\{\frac{\xi}{2} \int \!\!
    {\rm d}\omega_i {\rm d}\omega_s 
    \left[
    F_1(\omega_i, \omega_s) 
    {\rm e}^{2{\rm i} \phi}\,\,\hat a_1^\dagger(\omega_i)\hat a_1^\dagger(\omega_s)
    +F_2(\omega_i, \omega_s) \,\,\hat a_2^\dagger(\omega_i)\hat a_2^\dagger(\omega_s) \right]
    \right\}
    | {\rm vac} \rangle \,.
\end{align}

At this point, in \textbf{ Sec. IV}, the interference of the two routes $1$ and $2$ through a 2X2 MMI is discussed, whose behaviour on its two inputs may be analytically defined as follows:
\begin{align}
    \hat a_1^\dagger \to \frac{1}{\sqrt{2}} \left(\hat a_1^\dagger + {\rm i}\, \hat a_2^\dagger\right)
    \hspace{2cm}
    \hat a_2^\dagger \to \frac{1}{\sqrt{2}} \left(  \hat a_2^\dagger+{\rm i}\, \hat a_1^\dagger\right) \,.
\end{align}
Applying these transformations, the state will evolve as follows
\begin{align}
    |\Psi\rangle_{\rm IV}
    & = 
    \exp\Big\{ \frac{\xi}{4} \int \!\!
    {\rm d}\omega_i {\rm d}\omega_s 
    \Big[
    F_1(\omega_i, \omega_s) \,
    {\rm e}^{2{\rm i} \phi}\,
    \left(\hat a_1^\dagger(\omega_i)+{\rm i}\,\hat a_2^\dagger(\omega_i) \right) 
    \left( \hat a_1^\dagger(\omega_s)+{\rm i}\,\hat a_2^\dagger(\omega_s) \right) 
    \\
    &\hspace{2.5cm}
    +F_2(\omega_i, \omega_s) \,    \left(\hat a_2^\dagger(\omega_i)+{\rm i}\,\hat a_1^\dagger(\omega_i) \right) 
    \left( \hat a_2^\dagger(\omega_s)+{\rm i}\,\hat a_1^\dagger(\omega_s) \right)  \Big]
    \Big\}
    | {\rm vac} \rangle 
    \nonumber\\
    & = 
    \exp\Big\{ \frac{\xi}{4} \int \!\!
    {\rm d}\omega_i {\rm d}\omega_s
    \Big[
    \left(F_1(\omega_i, \omega_s) \,{\rm e}^{2{\rm i} \phi} - F_2(\omega_i, \omega_s) \right) 
    \left( \hat a_1^\dagger(\omega_i)\,\hat a_1^\dagger(\omega_s) -\hat a_2^\dagger(\omega_i)\,\hat a_2^\dagger(\omega_s) \right) 
    \nonumber\\
    &\hspace{2.5cm}
    + 2{\rm i}\,\left(F_1(\omega_i, \omega_s) \,{\rm e}^{2{\rm i} \phi} + F_2(\omega_i, \omega_s) \right) \hat a_1^\dagger(\omega_i)\,\hat a_2^\dagger(\omega_s) 
    \Big]
    \Big\}
    | {\rm vac} \rangle \,.
    \nonumber
\end{align}

The filtering and detection is represented by the operators
\begin{align}
    \hat P_i &:= \int \!\! {\rm d}\omega f(\omega) \sum_{n\ne 0} \frac{1}{n!} \left(\hat a_i(\omega)^\dagger\right)^n | {\rm vac} \rangle
    \langle {\rm vac}| \left(\hat a_i(\omega)\right)^n
    \label{eq:pro_op}
\end{align}
where the function $f$ represents the spectral amplitude of the filter.
Note that we have implemented operators that take into account threshold detectors.

The probability of detecting coincidences in channels $1$ and $2$ (Fig. \ref{fig:SiPIC-1}(a)) is given by
\begin{align}
    p_{12}&= \frac{{\rm Tr} \left\{ |\psi^{\rm out}\rangle \langle \psi^{\rm out}| \hat P_1 \otimes \hat P_2  \right\} }{{\rm Tr} \left\{ |\psi^{\rm out}\rangle \langle \psi^{\rm out}|  \right\}}
    = \frac{\langle \psi^{\rm out}| \hat P_1 \otimes \hat P_2 |\psi^{\rm out}\rangle}{\langle \psi^{\rm out} |\psi^{\rm out}\rangle}
    \nonumber\\
    & \approx \frac{1}{4} \int \!\!
     {\rm d}\omega_i {\rm d}\omega_s 
     f(\omega_i) f(\omega_s) 
    \Big[
     | F_1(\omega_i, \omega_s)|^2
    +| F_2(\omega_i, \omega_s)|^2 
    +2 {\rm Re} \Big[ F_1(\omega_i, \omega_s) \bar F_2(\omega_i, \omega_s)
    {\rm e}^{2{\rm i} \phi} \Big]
    \Big] 
    \nonumber\\
    &=\frac{1}{2} \left\{ 1 + 
    {\rm Re}\Big[
    {\rm e}^{2{\rm i} \phi}
    \int \!\! {\rm d}\omega_i {\rm d}\omega_s
    f(\omega_i) f(\omega_s) 
    F_1(\omega_i, \omega_s) \bar F_2(\omega_i, \omega_s)
    \Big] \right\} \,.
    \label{theo_coinc_sep}
\end{align}

The approximation consisting of keeping only the leading term is based on the fact that we can choose to work in the low gain regime once the sources have been characterized. 

We conclude this subsection, by writing the visibility in terms of the JSAs overlap. First, the visibility $V$ is defined as \cite{branczyk2017hong}
\begin{align}
    V = \frac{p_{12}^{\rm max} - p_{12}^{\rm min}}{p_{12}^{\rm max}} \,.
    \label{V_prob}
\end{align}
This definition has been chosen in the analysis of the experimental data, but there are other choices \cite{Adcock_2021}, which lead to different values of visibility. Nevertheless, in our cases the visibility can be linked to a physical quantity that is independent from any convention. 

In terms of JSAs overlap, $p_{12}$ takes the form
\begin{align}
    p_{12}(\phi, \delta, N ) &= \frac{1}{2} \left\{ 1 + 
    N \cos\left(2 \phi + \delta \right)
     \right\} \,,
     \\
     \mbox{where}\quad & N\, {\rm e}^{{\rm i} \delta} := \int \!\! {\rm d}\omega_i {\rm d}\omega_s
    f(\omega_i) f(\omega_s) 
    F_1(\omega_i, \omega_s) \bar F_2(\omega_i, \omega_s) \,.
\end{align}
It is easy to see that $F_1 = F_2$ implies $N=1$ and $\delta=0$ and, therefore,  $p_{12}= \cos(\phi)^2$ and $p_{12}^{\rm max} = p_{12}(\phi = 0)$ and $p_{12}^{\rm min} = p_{12}(\phi = \pi/2)$ . \\
In the generic case, the maxima of $p_{12}$ are located at $2 \phi + \delta = 2 N \pi$ and the minima at $2\phi + \delta = N \pi$ with $N \in \mathbf{Z}$ and the visibility reads
\begin{align}
    V &= \frac{2N}{1+N}\,,
     \\
     \mbox{where}\quad & N := \Big|\int \!\! {\rm d}\omega_i {\rm d}\omega_s
    f(\omega_i) f(\omega_s) 
    F_1(\omega_i, \omega_s) \bar F_2(\omega_i, \omega_s) \Big| \,.
    \label{eq:vis_vs_overlap}
\end{align}

Therefore, knowing the experimental value for V, it is possible to deduce N as we did in Table \ref{tab:summary} of the main text where the computed JSAs overlaps for our sources are reported.

The normalized coincidence probability reads
\begin{align}
    p_{12}^{\rm norm}(\phi, \delta, N) &= \frac{1}{1+N} \left\{ 1 + 
    N \cos\left(2 \phi + \delta \right)
     \right\} \,.
\end{align}
This formula has been used to build the theoretical results in Figs. \ref{fig:revHOM_MZI_spiral}(b), \ref{fig:HOM_Ring_and_Spiral}(b)-(c) and \ref{fig:MZIringspiral_HOM}.\\
From the formula for the normalized coincidence probability we can note that a relative phase difference between the JSA of the two sources can shift the interference pattern of the HOM experiment.
Of course, such relative phase can be easily hidden from our observation because of non-idealities due to the fabrication: in our case a small difference in the length of the two arms of the MZI involved in the HOM dynamics should give a big contribution.

\subsection{Evolution state description in the reverse HOM-HOM-reverse HOM experiment}
\label{app:HOM-RevHOM}

In this section, we examine the theoretical underpinnings \cite{hong1987measurement, branczyk2017hong, bouchard2020two} of the reverse HOM-HOM-reverse HOM experiment as seen in the integrated system shown in Figs. \ref{fig:SiPIC-1}(c) and \ref{fig:SiPIC-2}(a). 

As we have done in the previous subsection, we divide the integrated system in different sections. Note that by changing $\phi$ to $\phi_1$, the design is the same until \textbf{Sec. IV}.
In \textbf{ Sec. V}, the two paths are divided into other two paths through two 1X2 MMIs. At this step, path $1$ is split into two new paths, which will be denoted as path $1$ (corresponding to detection channel $1$) and $3$ (corresponding to detection channel $3$), and analogously path $2$ into path $4$ (corresponding to detection channel $4$) and path $2$ (corresponding to detection channel $2$). In \textbf{Sec. VI}, an integrated MZI made up of two 2X2 MMIs and a phase shifter causes photons from paths $3$ and $4$ to interact. The filtering and detection of the four pathways are carried out in the final \textbf{Sec. VII}.
\begin{figure}[http]
    \centering
    \includegraphics[scale=0.4]{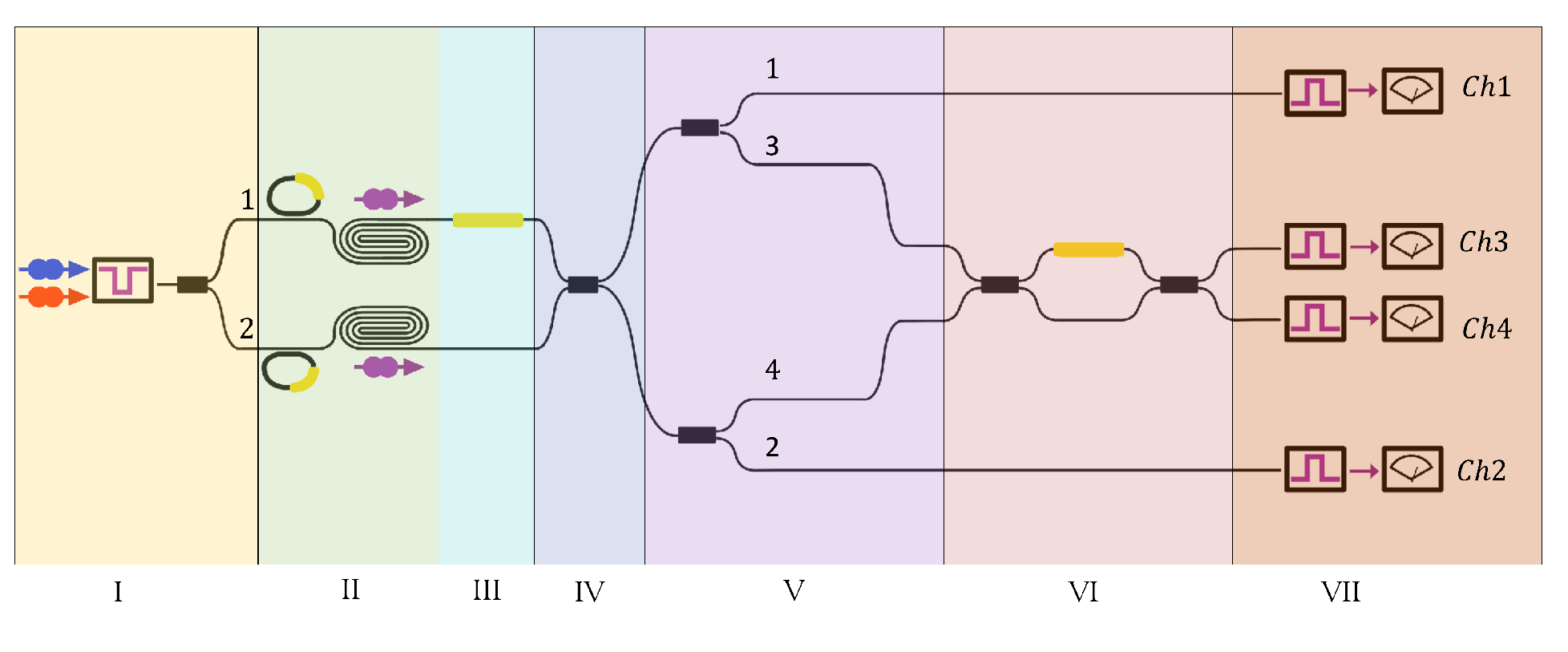}

\caption{\small{ The chip design shown in Fig. \ref{fig:SiPIC-1}(c) 
is displayed. It is divided into 7 parts for simplicity of discussion. }
}
    \label{design}
\end{figure}
\\
Let's start from the state in \textbf{ Sec. IV}
\begin{align}
    |\Psi\rangle_{\rm IV}
    & = 
    \exp\Big\{ \frac{\xi}{4} \int \!\!
    {\rm d}\omega_i {\rm d}\omega_s 
    \Big[
    F_1(\omega_i, \omega_s) \,
    {\rm e}^{2{\rm i} \phi}\,
    \left(\hat a_1^\dagger(\omega_i)+{\rm i}\,\hat a_2^\dagger(\omega_i) \right) 
    \left( \hat a_1^\dagger(\omega_s)+{\rm i}\,\hat a_2^\dagger(\omega_s) \right) 
    \\
    &\hspace{2cm}
    +F_2(\omega_i, \omega_s) \,    \left(\hat a_2^\dagger(\omega_i)+{\rm i}\,\hat a_1^\dagger(\omega_i) \right) 
    \left( \hat a_2^\dagger(\omega_s)+{\rm i}\,\hat a_1^\dagger(\omega_s) \right)  \Big]
    \Big\}
    | {\rm vac} \rangle \,.
    \nonumber
\end{align}

In \textbf{Sec. V}, we have two 1x2 MMI described as:
\begin{align}
    \hat a_1^\dagger \to \frac{1}{\sqrt{2}} \left(\hat a_1^\dagger + \hat a_3^\dagger\right)
    \hspace{2cm}
    \hat a_2^\dagger \to \frac{1}{\sqrt{2}} \left( \hat a_4^\dagger + \hat a_2^\dagger\right)
\end{align}
where with an abuse of notation $\hat a_1^\dagger$ corresponds to the external upper waveguide, $\hat a_3^\dagger$ to the central upper one, $\hat a_4^\dagger$ to the central lower one and $\hat a_2^\dagger$ to the external lower one (see the corresponding detection channels Fig. \ref{design}).
Thus, at \textbf{Sec. V} the state will be
\begin{align}
    |\Psi\rangle_{\rm V}
    & = 
    \exp\Big\{ \frac{\xi}{8} \int \!\!
    {\rm d}\omega_i {\rm d}\omega_s
    \Big[
    F_1(\omega_i, \omega_s) \,
    {\rm e}^{2{\rm i} \phi_1}\,
    \hat b_1^\dagger(\omega_i) \, \hat b_1^\dagger(\omega_s) 
    -\,F_2(\omega_i, \omega_s) \,   
    \hat b_2^\dagger(\omega_i) \, \hat b_2^\dagger(\omega_s) 
    \Big\}
    | {\rm vac} \rangle \,,
    \nonumber\\
    & {\rm where} \quad 
    \hat b_1^\dagger(\omega) \equiv \hat a_1^\dagger(\omega)+\hat a_3^\dagger(\omega)+{\rm i}\,\hat a_4^\dagger(\omega)+{\rm i}\,\hat a_2^\dagger(\omega) \,,
    \nonumber\\
    & {\rm and} \quad\quad
    \hat b_2^\dagger(\omega) \equiv 
    \hat a_1^\dagger(\omega)+\hat a_3^\dagger(\omega)-{\rm i}\,\hat a_4^\dagger(\omega)-{\rm i}\,\hat a_2^\dagger(\omega) \,.
    \nonumber
\end{align}
The second MZI in \textbf{Sec. VI} solely acts on photons traveling along pathways $\hat a_3$ and $\hat a_4$. 
Therefore, the resulting state in \textbf{Sec. VI} reads
\begin{align}
    |\Psi\rangle_{\rm VI}
    & = 
    \exp\Big\{ \frac{\xi}{8} \int \!\!
    {\rm d}\omega_i {\rm d}\omega_s
    \Big[
    F_1(\omega_i, \omega_s) \,
    {\rm e}^{2{\rm i} \phi_1}\,
    \hat b_3^\dagger(\omega_i) \, \hat b_3^\dagger(\omega_s) 
    -\,F_2(\omega_i, \omega_s) \,  
    \hat b_4^\dagger(\omega_i) \, \hat b_4^\dagger(\omega_s) 
    \Big\}
    | {\rm vac} \rangle \,,
    \nonumber\\
    & {\rm where} \quad 
    \hat b_3^\dagger(\omega) \equiv \hat a_1^\dagger(\omega)-\hat a_3^\dagger(\omega)+{\rm i}\,\hat a_4^\dagger(\omega)+{\rm i}\,\hat a_2^\dagger(\omega) \,,
    \nonumber\\
    & {\rm and} \quad\quad
    \hat b_4^\dagger(\omega) \equiv 
    \hat a_1^\dagger(\omega)
    +{\rm e}^{{\rm i}\phi_2}\left( \hat a_3^\dagger(\omega)+{\rm i}\,\hat a_4^\dagger(\omega) \right)-{\rm i}\,\hat a_2^\dagger(\omega) \,.
    \nonumber
\end{align}
The final state for the scheme shown in Fig. \ref{fig:SiPIC-2}(a) can be found tracing out the photons in channels 1 and 2, since in that case the 1x2 MMIs are substituted by asymmetric MZIs for pumps removal that do not affect the generated photons. Between schemes in Figs. \ref{fig:SiPIC-1}(c) and \ref{fig:SiPIC-2}(a), the formula for the visibility does not change as long as we replace ch3 and ch4 of Fig. \ref{fig:SiPIC-1}(c) with ch1 and ch2 of Fig. \ref{fig:SiPIC-2}(a), respectively.

The filtering and detection is represented by the operators defined in \eqref{eq:pro_op},
where the function $f$ represents the spectral amplitude of the filter.

The probability of detecting in channels $i$ and $j$ is given by
\begin{align}
    p_{ij}&= \frac{{\rm Tr} \left\{ |\psi^{\rm out}\rangle \langle \psi^{\rm out}| \hat P_i \otimes \hat P_j  \right\} }{{\rm Tr} \left\{ |\psi^{\rm out}\rangle \langle \psi^{\rm out}|  \right\}}
    = \frac{\langle \psi^{\rm out}| \hat P_i \otimes \hat P_j |\psi^{\rm out}\rangle}{\langle \psi^{\rm out} |\psi^{\rm out}\rangle} \,.
\end{align}

If we consider the probability after \textbf{Sec. VII}
\begin{align}
    p_{13}=p_{24}
    &= \frac{1}{8} \int \!\!
    \!
     {\rm d}\omega_i {\rm d}\omega_s 
     f(\omega_i) f(\omega_s) 
    \nonumber\\
    &\times
    \Big[
     | F_1(\omega_i, \omega_s)|^2
    +| F_2(\omega_i, \omega_s)|^2 +
    2\,{\rm Re}\Big[ F_1(\omega_i, \omega_s) \bar F_2(\omega_i, \omega_s)
    {\rm e}^{{\rm i} (2\phi_1-\phi_2)} \Big] 
     \Big] 
    \nonumber\\
    p_{14}=p_{23}
    &= \frac{1}{8} \int \!\!
    \!
     {\rm d}\omega_i {\rm d}\omega_s 
     f(\omega_i) f(\omega_s) 
    \nonumber\\
    &\times
    \Big[
     | F_1(\omega_i, \omega_s)|^2
    +| F_2(\omega_i, \omega_s)|^2 -
    2\,{\rm Re}\Big[ F_1(\omega_i, \omega_s) \bar F_2(\omega_i, \omega_s)
    {\rm e}^{{\rm i} (2\phi_1-\phi_2)} \Big] 
    \Big] 
    \nonumber\\
    p_{12}
    &= \frac{1}{8} \int \!\!
    \!
     {\rm d}\omega_i {\rm d}\omega_s 
     f(\omega_i) f(\omega_s) 
    \nonumber\\
    &\times
    \Big[
     | F_1(\omega_i, \omega_s)|^2
    +| F_2(\omega_i, \omega_s)|^2 +
    2\,{\rm Re}\Big[ F_1(\omega_i, \omega_s) \bar F_2(\omega_i, \omega_s)
    {\rm e}^{2{\rm i} \phi_1}\Big]
    \Big] 
    \nonumber\\
    p_{34}
    &= \frac{1}{8} \int \!\!
    \!
     {\rm d}\omega_i {\rm d}\omega_s 
     f(\omega_i) f(\omega_s) 
    \nonumber\\
    &\times
    \Big[
     | F_1(\omega_i, \omega_s)|^2
    +| F_2(\omega_i, \omega_s)|^2 +
    2\,{\rm Re}\Big[F_1(\omega_i, \omega_s) \bar F_2(\omega_i, \omega_s)
    {\rm e}^{2{\rm i} (\phi_1-\phi_2)} \Big]
    \Big] 
    \nonumber
\end{align}
where the 4 detectors are represented by the indices 1, 2, 3, and 4. Detectors 1 and 2 are specifically the outside detectors that correspond to pathways $a_1$ and $a_2$. Detectors 3 and 4 make up the two central detectors corresponding to the two paths $a_3$ and $a_4$.
We can rewrite $p_{34}$ and $p_{12}$ as
\begin{align}
    p_{12}
    &= \frac{1}{8} \left\{ 1+ 
    {\rm Re}\Big[
    {\rm e}^{2{\rm i} \phi_1}
    \int \!\! {\rm d}\omega_i {\rm d}\omega_s
    f(\omega_i) f(\omega_s) 
    F_1(\omega_i, \omega_s) \bar F_2(\omega_i, \omega_s)
    \Big] \right\}\,,
    \\
    p_{34}
    &= \frac{1}{8} \left\{ 1 + 
    {\rm Re}\Big[
    {\rm e}^{2{\rm i} (\phi_1-\phi_2)}
    \int \!\! {\rm d}\omega_i {\rm d}\omega_s
    f(\omega_i) f(\omega_s) 
    F_1(\omega_i, \omega_s) \bar F_2(\omega_i, \omega_s)
    \Big] \right\} \,.
\end{align}

The visibility is defined as 
\begin{align}
    V = \frac{p_{34}^{\rm max} - p_{34}^{\rm min}}{p_{34}^{\rm max}}
    \label{V_prob_bis}
\end{align}
and doing the same manipulation as in \ref{app:RevHOM} we can arrive at the same formulas.
The same conclusion is due to the fact that the all the designs exploit the same features once the data of the coincidences are normalized.

\subsection{Impact of pump noise and fabrication imperfections on the visibility and JSAs overlap}
\label{noise_intro}

In this section, we want to investigate how the presence of unwanted pump noise can affect the visibility of the interference pattern of coincidences measurements.
We model such noise as an additional contribution to the density matrix given by the desired output state $|\psi_{\rm out}\rangle$ obtained at the end of the manipulation done on the PIC.
The total density matrix reads
\begin{align}
    \rho_{\rm tot} = c \,|\psi_{\rm out}\rangle\langle\psi_{\rm out}| + (1-c) \,\rho_{\rm noise}\,,
\end{align}
where $c$ is a visibility parameter that depends on the signal to noise ratio, CAR and the experimental setup, and $\rho_{\rm noise}$ represents the noise contribution at the output.
The probability of detecting in channels $i$ and $j$ is modified and has the following form
\begin{align}
    p_{ij}&= \frac{{\rm Tr} \left\{ \rho_{\rm tot} \hat P_i \otimes \hat P_j  \right\} }{{\rm Tr} \left\{ \rho_{\rm tot}  \right\}}
    = \frac{c\,\langle \psi^{\rm out}| \hat P_i \otimes \hat P_j |\psi^{\rm out}\rangle + (1-c)\,{\rm Tr} \left\{ \rho_{\rm noise} \hat P_i \otimes \hat P_j  \right\} }{{\rm Tr} \left\{ \rho_{\rm tot}  \right\}}  \,.
\end{align}
Using the definition of visibility we proposed in this work and the results of the previous subsections, we obtain
\begin{align}
    V &= \frac{2N}{1+N + D}\,,
\end{align}
where $N$ is the JSAs overlap given in Eq. \eqref{eq:vis_vs_overlap} and $D$ is a parameter constructed from ${\rm Tr} \left\{ \rho_{\rm noise} \hat P_i \otimes \hat P_j  \right\}$ and $c$.
The previous equation is obtained assuming that the noisy part depends weakly on the $\phi$s phases and it shows qualitatively lower visibility for higher noise, i.e. $c\to 0$.
In Fig. \ref{fig:MZIringspiral_HOM} of Sec. \ref{sec:ring} we show how the visibility increases thanks to a better filtering of the residual pump through an aMZI and a consequent better CAR value.
For future investigations it would be interesting to link the parameter $D$ to the CAR and quantitatively estimate its contribution. But this goes beyond the scope of this paper and is left for further studies.

\begin{figure}[http]
    \centering
    a)\includegraphics[scale=0.28]{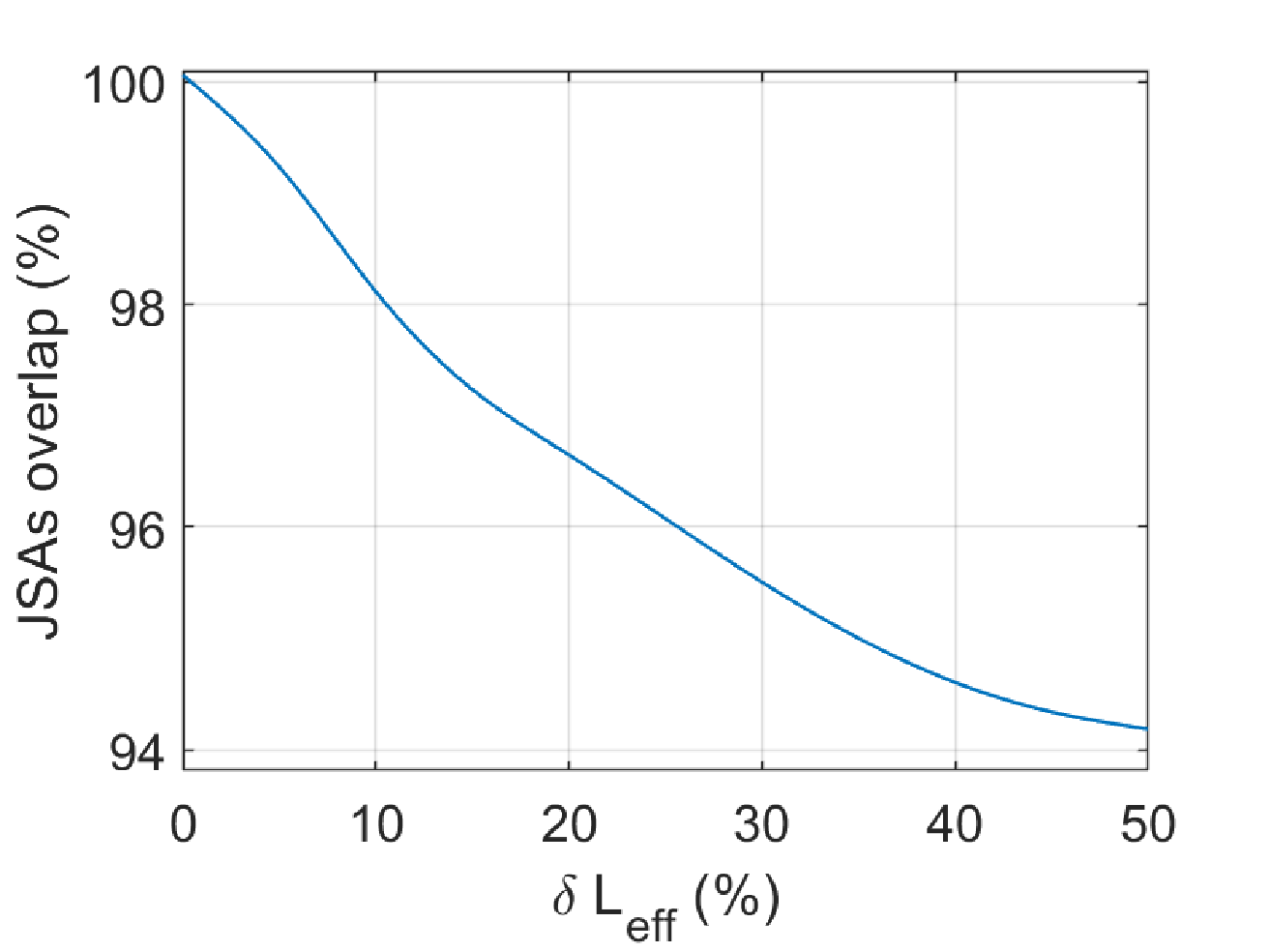}b)
    \includegraphics[scale=0.28]{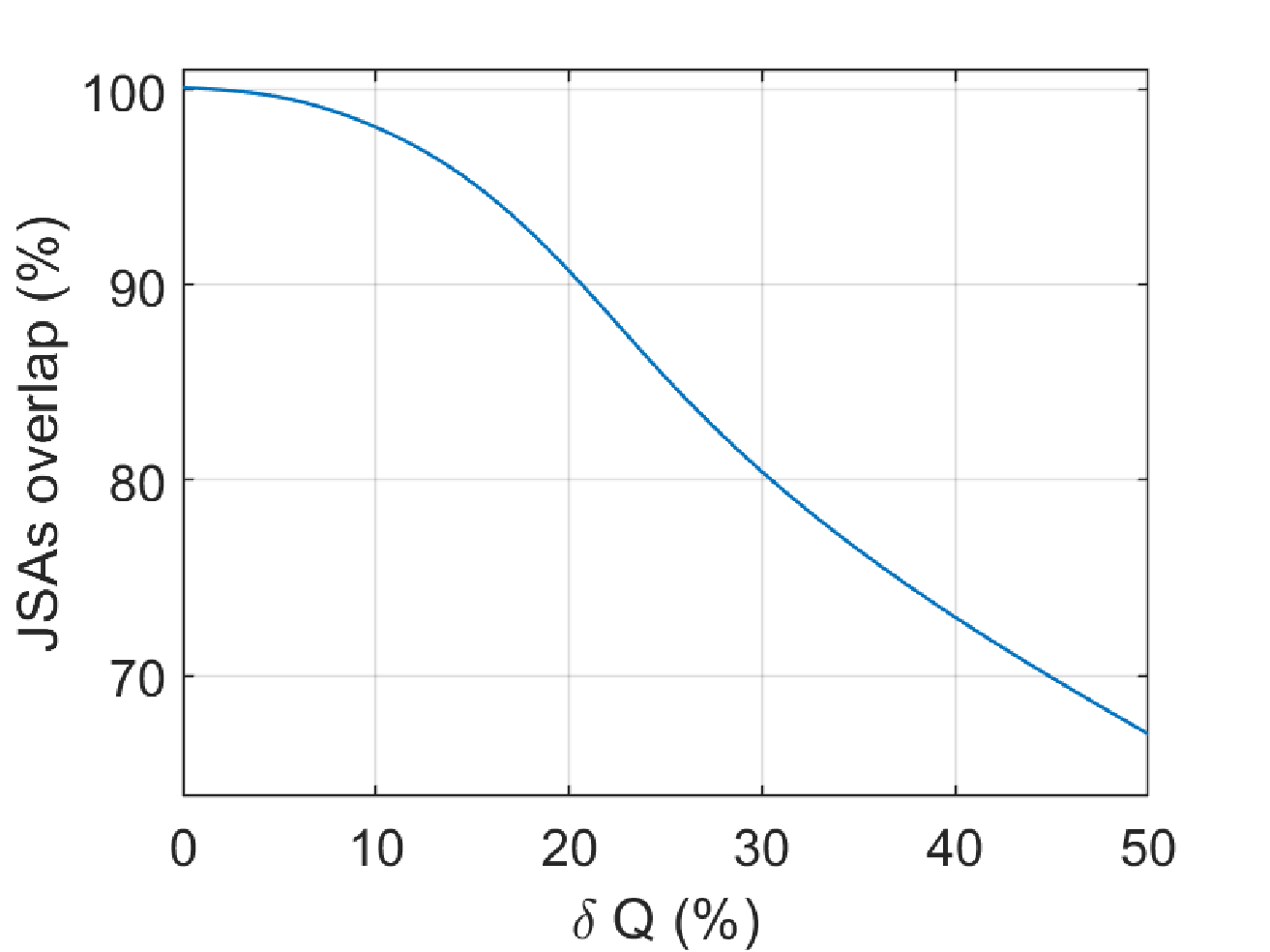}
\caption{\small{ Simulated JSAs overlap as a function of variations of $L_{\rm eff}$ for spiral waveguides (a) and JSAs overlap as a function of variations of Q-factor for microring resonators (b). }
}
    \label{simu_imp}
\end{figure}

Finally, in Fig. \ref{simu_imp} we report the plots showing the behavior of the JSAs overlap assuming deviations in the parameters of one of the two sources. In particular, for spiral waveguides we consider variations of $L_{\rm eff}$ ($\delta L_{\rm eff} = \Delta L_{\rm eff}^{\rm error}/L_{\rm eff}^{\rm nominal} [\%]$), while for microring resonators we consider variations of the Q-factor ($\delta Q= \Delta Q^{\rm error}/Q^{\rm nominal} [\%]$).



\bibliography{HOM.bib}






\end{document}